\documentclass{aa}  

\usepackage{graphicx}
\usepackage{natbib}
\usepackage{newtxtext}
\usepackage{booktabs}
\usepackage{longtable}
\usepackage{txfonts}
\usepackage{subcaption}
\usepackage[T1]{fontenc}
\usepackage{ae,aecompl}
\usepackage{lscape}
\usepackage{placeins}
\usepackage{float}
\usepackage{multicol}
\bibpunct{(}{)}{;}{a}{}{,}
\usepackage{xkeyval}
\usepackage{hyperref}
\hypersetup{hidelinks}

\newcommand{\chisq}{$\chi^{2}$}
\newcommand{\ironab}{$\left[\mathrm{Fe} / \mathrm{H}\right]$\,}

\newcommand{\Gbp}{$\mathrm{G}_{\mathrm{BP}}$ \,}
\newcommand{\Grp}{$\mathrm{G}_{\mathrm{RP}}$ \,}

\newcommand{\Rsun}{$R_{\odot}$\,}
\newcommand{\logg}{$\log{g}$\,}

\newcommand{\teffnospace}{$T_{\rm eff}$}
\newcommand{\loggnospace}{$\log{g}$}
\newcommand{\vmicnospace}{$\mathrm{V}_{\mathrm{mic}}$}
\newcommand{\vbrdnospace}{$\rm{V}_{\rm{brd}}$}

\begin{document}

\title{Classical Cepheids in the 4MOST era}
\subtitle{I. Methodology for 4MOST variable stars and NLTE abundance gradients in the Milky Way.}
\titlerunning{Classical Cepheids in the 4MOST era I}

   \author{R. Albarracín\inst{1, 2}\fnmsep\thanks{Email: rialbarracin@mpia.de}
        \and M. Bergemann\inst{1}
        \and N. Storm\inst{1,2}
        \and G. Guiglion\inst{1}
        \and R.~I. Anderson\inst{3}
        \and V. D’Orazi \inst{4, 5}
        \and A. Dimoff\inst{1}
        \and G. Kordopatis\inst{6}
        \and G. Bono\inst{4, 5}
        \and T. Bensby\inst{7}
        \and G. Traven \inst{8}
        \and R. Church \inst{7}
        }

    \institute{
    Max-Planck-Institut f\"{u}r Astronomie, K\"{o}nigstuhl 17, D-69117 Heidelberg, Germany
    \and
    Heidelberg University, Grabengasse 1, 69117 Heidelberg, Germany
    \and
    Institut f\"ur Astrophysik und Geophysik, Georg-August-Universit\"at G\"ottingen, Friedrich-Hund-Platz 1, 37077 G\"ottingen, Germany
    \and
    Dept. of Physics, University of Rome Tor Vergata, via della Ricerca Scientifica 1, 00133, Rome, Italy
    \and 
    INAF Astronomical Observatory of Rome, via Frascati 33, 00078, Monte Porzio Catone (RM), Italy
    \and
    Universit\'e C\^ote d’Azur, Observatoire de la C\^ote d’Azur, CNRS, Laboratoire Lagrange, Nice, France
    \and 
    Lund Observatory, Division of Astrophysics, Department of Physics, Lund University, Box 118, SE-221\,00 Lund, Sweden
    \and
    Faculty of Mathematics and Physics, University of Ljubljana, Jadranska 19, 1000 Ljubljana, Slovenia
    }

   \date{Received 30 July 2026 / Accepted 2 November 2026}
   
  \abstract
   {Classical Cepheids (CCs) are key distance indicators and tracers of young stellar populations in the Milky Way. Their period-luminosity-metallicity (PLZ) relation also makes them calibrators of the Hubble constant $H_0$. Despite their importance, chemical composition, especially non-LTE (NLTE) effects, remains a significant source of systematics in PLZ applications.}
   {In preparation for the large sample of CCs expected from the 4MIDABLE-HR (4MOST Consortium Survey 4: Milky Way Disc and Bulge High-Resolution) survey, we present a method for deriving Cepheid atmospheric parameters and chemical abundances in NLTE.} 
   {Using the Payne artificial neural network (ANN) trained on NLTE spectra, we derive atmospheric parameters and abundances for 17 elements: O, Na, Mg, Al, Si, Ca, Ti, Mn, Fe, Co, Ni, Sr, Y, Ba, Eu in NLTE and C and Cr in LTE. We apply the new method to archival spectra of Cepheids and derive radial chemical gradients in the Milky Way.}
   {We derive robust NLTE stellar parameters and abundances for Cepheids spanning a broad metallicity range, from $\sim -0.87 \, \rm{dex}$ to $\sim +0.35 \, \rm{dex}$. For stars with multi-epoch spectra, we find that NLTE abundances are consistent across the pulsation cycle, whereas LTE values show larger dispersions and biases. The resulting abundance distributions and gradients align well with Galactic disk expectations from chemical evolution models. We confirm that the gradients flatten in the outer disk beyond $R_{\rm GC} \sim 15\,\mathrm{kpc}$ and provide the first NLTE gradients for Co, Ni, Y, and Eu.}
   {We show that our NLTE astrophysical analysis yields robust solutions for atmospheric parameters and chemical abundances across spectra at the resolving power of 4MOST-HR. This enables large, homogeneous studies with upcoming surveys such as 4MOST 4MIDABLE-HR and provides a solid basis for tightening constraints on Galactic chemical evolution models and Cepheid distance-scale work.}


   \maketitle
\nolinenumbers

\section{Introduction}

Classical Cepheids (CCs) are young, intermediate-mass (3-12 $M_\odot$) core He-burning stars \citep{Turner1996, Bono2000b, Bono2024} residing in the instability strip region of the Hertzsprung–Russell diagram. Typical ages for CCs can span from 30 Myr to 600 Myr depending on the mass and chemical composition \citep{Bono05, Anderson16}, and are therefore found in Galaxies with recent star formation. These stars exhibit radial pulsations with typical periods of a few days, although their periods can extend beyond 100 days \citep{Marciobook}. They have luminosities of $\sim 10^{4}-10^{5} L_{\odot}$ and radii of 20-200 $R_{\odot}$ \citep{Anderson2014}. During their evolution, CC progenitors may cross the instability strip multiple times and experience dredge-up processes, shaping their observed pulsation properties and surface chemical abundances \citep{Anderson2014, Anderson16}.

Owing to the tight relation between pulsation period and intrinsic luminosity, CCs are among the first and best-established standard candles \citep{LeavittLaw1, LeavittLaw2, Fernie1969, SandageTamman06, Riess2018, Ripepi2026}. Due to their intrinsic brightness, CCs have canonically been used to calibrate the extragalactic distance scale \citep{MadoreFreedman91, Udalski99, Benedict07, Riess2021} and thus to determine the Hubble constant \citep{Freedman2012, Riess2016, Riess2019, Riess2022}. Additionally, CCs are excellent tracers of structure and chemical evolution of young stellar populations in the Milky Way \citep{lemasle2017, Skowron2019} and other star-forming galaxies \citep{Viradell2007, Neeley21, DeSomma2025}. In contrast to older stars affected by radial migration \citep{Minchev2013, Lehman2024, Kordopatis25}, CCs are more direct proxies of the composition of the present-day ISM \citep{Lemasle13}, thereby providing constraints on models of Galactic Chemical Evolution (GCE) and parameters such as the efficiency of star formation \citep{Palla2020, Matteucci2021Rev, Lian2023, Storm2026}. Moreover, since CCs are spread across all four Galactic quadrants and are closely related with the spiral arms \citep{Lemasle2022}, they are of major use in the analyses of non-axisymmetric evolution \citep{Spitoni2023}, linked to the dynamics of the spiral arms and the bar \citep{Fragkoudi2018, Hackshaw2024, Debattista2025}. 

The chemical composition of CCs affects their pulsation properties through its impact on envelope opacity and, therefore, on the efficiency of the $\kappa$-mechanism in the H and He ionization regions \citep{Bono99, Marconi05}. In practice, a higher metal content, $Z$, increases the envelope opacity and convective efficiency, reducing pulsation driving in the hydrogen ionization regions \citep{Bono99, Caputo00}. This shifts the instability strip toward lower effective temperatures, that is, to redder colors at fixed stellar mass \citep{Anderson16, DeSomma2022, Deka2024}. 

As a result, metallicity changes the topology of the instability strip and affects the calibration of the Cepheid period–luminosity (PL) relation \citep{Sakai2004, Fiorentino07}, motivating the use of a period–luminosity–metallicity (PLZ) relation. However, the sign and magnitude of the metallicity effect remain highly debated. Some observational studies find that solar-metallicity Cepheids are brighter at fixed period than metal-poor counterparts \citep{Breuval22, Breuval2024, Trentin24a}, whereas other observations \citep{Romaniello08} and pulsation models \citep{ Bono99-2, Caputo00, Marconi05} suggest the opposite trend. For these reasons, accurate stellar parameters and chemical abundances are essential for the modeling of pulsations in CCs, and their use in the extragalactic distance-scale and stellar population studies \citep{SandageTamman06, Storm2011, Riess2021, Ripepi2022}.

In this work, we introduce an astrophysical analysis method to analyze spectra of Cepheids to be observed within the 4MIDABLE-HR survey \citep[4MOST Consortium Survey 4: Milky Way Disc and Bulge High-Resolution;][]{Bensby2019, S4paper}. In the context of this survey, Cepheids will be used to explore the detailed structure of the MW disk, and through a collaborative agreement with a supplementary science program (lead V. Ripepi, see \citet{S4paper}), to investigate the role of composition in the PLZ studies. The 4MIDABLE-HR survey \citep{Bensby2019, S4paper} will provide high-resolution ($R \sim 20{,}000$) optical spectra in three windows: $3926 - 4355$\,\AA, $5160 - 5730$\,\AA, and $6100 - 6790$\,\AA \ \citep{4MOST_dejong}, of several million stars in the Milky Way disk and bulge. Its dedicated Cepheid sub-survey will collect high-quality spectra with the SNR of $\gtrsim 100$ per \AA~for $2\,643$ Cepheids with $G\leq16.5$, covering the Galactic bulge and disk. Among these variable stars, 1423 are CCs, 90 are Anomalous Cepheids, 270 dwarf Cepheids, and 860 are type 2 Cepheids. This spectroscopic dataset will enable the determination of precise and homogeneous abundances of all key nucleosynthesis channels across the disk, including elements such as C, O, Na, Mg, Al, Si, Ca, Ti, Mn, Co, Ni, Sr, Y, Ba, Eu. This can be used for a detailed analysis of the effects of metallicity in the PLZ relation, tracing the present-day chemistry of the Milky Way, and the evolution of intermediate-mass stars.

The main novelty of our approach is the use of non-local thermodynamic equilibrium (NLTE) line formation for 15 elements. This is particularly important for CCs, whose warm, low-density, and extended atmospheres favor departures from LTE \citep[see][for a review]{Bergemann2025}. Under these conditions, the radiation field can deviate strongly from the local thermal field, and radiative rates may exceed collisional rates by several orders of magnitude. As a result, NLTE effects can significantly alter line profiles and, consequently, the inferred chemical abundances. Most previous abundance studies of CCs have relied on LTE spectral synthesis \citep[e.g.][]{Kovtyukh00, Romaniello08, Luck2011, Luck13, Lemasle13, Genovali14, Genovali15, DaSilva16, Romaniello22, DaSilva23, Trentin23, Trentin24b}, or have applied NLTE corrections to selected elements \citep{Andrievsky2013, Takeda2013}. Recently, \citet{Nunnari25} performed a comprehensive NLTE investigation of Cepheid abundances using the SME code \citep{SME, PySME}, finding significant abundance offsets for elements such as O of $\sim -0.7$ dex. These results highlight the need for a consistent NLTE treatment in Cepheid abundance analyses. In this work, we develop a Cepheid analysis method based on NLTE TurboSpectrum calculations. This choice ensures consistency with the NLTE TurboSpectrum grids that will be used by 4MIDABLE-HR and other 4MOST surveys \citep[TurboSpectrum v20;][]{Gerber2023, Storm2023, Storm2026}, and provides a homogeneous framework for deriving atmospheric parameters and chemical abundances for the 4MOST Cepheid products.

The paper is structured as follows. In Section \ref{sec:Data}, we describe the observed archival data to be used as validation of the spectroscopic analysis code. Section \ref{sec:methods} describes the NLTE spectrum synthesis, the Artificial Neural Network (ANN) based emulator, trained on the NLTE spectra for the analysis of atmospheric parameters and abundances of CCs and other variables in our validation sample. In Section \ref{sec:results}, we present the results of the analysis of the archival high-resolution data obtained with FEROS, UVES and HARPS for 415 stars in NLTE, and then discuss the NLTE abundances and radial chemical gradients in the context of GCE models. In Section \ref{sec:discussion} we discuss the results and compare with the previous literature work, and summarize the conclusions in Section \ref{sec:conclusions}.

\section{Data}
\label{sec:Data}

\begin{figure}
	\includegraphics[width=\columnwidth]{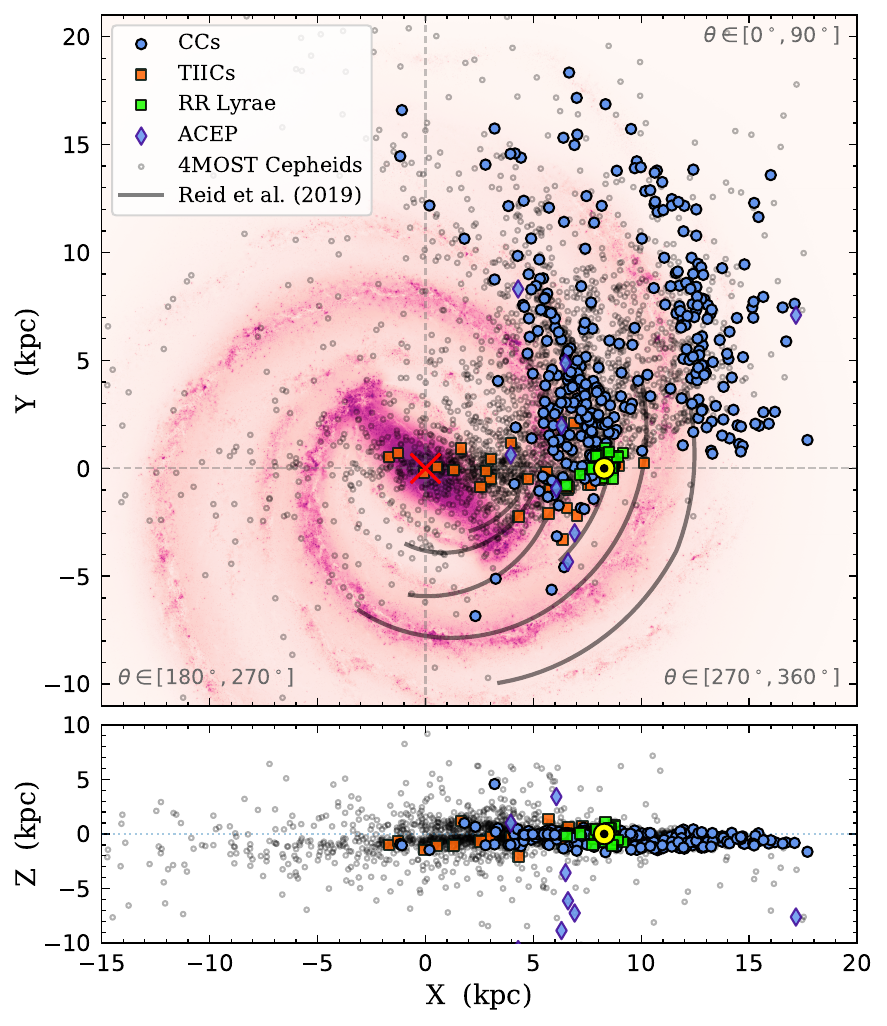}
    \caption{Three-dimensional distribution of Cepheids and other variables in this work in the MW. Blue dots, blue diamonds, orange squares, and light-green squares indicate CCs, Anomalous Cepheids, Type II Cepheids, and RR Lyrae stars, respectively, with available archival spectroscopic data. Gray open circles show the 4MIDABLE-HR input catalog of Cepheids. The Sun is located at $(X,Y)=(8.277,0)$~kpc \citep{GCdist} and is marked with a yellow dot, while the Galactic center is marked with a red cross at $(X,Y)=(0,0)$. Vertical and horizontal gray lines show the limits of the Galactic quadrants, with the range of the azimuthal coordinates in the respective corners. Log-periodic spiral-arm segment fits of the Local Arm, based on trigonometric parallaxes of high-mass star-forming regions from \citet{Reid2019}, are shown as solid black lines. The bottom panel shows the $XZ$ view of the distribution in the MW. The face-on map of the Milky Way was made using the Python library \texttt{mw-plot}.}
    \label{fig:MWFaceon}
\end{figure}

To validate the methodology that will be applied to the analysis of Cepheid spectra in the 4MOST 4MIDABLE-HR survey, we consider an initial sample of 1108 publicly available archival spectra, corresponding to 419 Classical Cepheids.

The archival spectra were obtained with high-resolution spectrographs at large facilities: UVES \citep{Dekker2000} at the VLT, FEROS \citep{Kaufer99} at La Silla, and HARPS \citep{Mayor2003} at the ESO 3.6-m telescope at La Silla, with resolving powers of $R\!\sim\!47\,000$, $R\!\sim\!48\,000$ and $R\!\sim\!115\,000$, respectively. The S/N of these data ranges from 21 to 412 per \AA. To ensure consistency with the upcoming 4MOST HR spectral data ($R\!\sim\!18\,000$--$21\,000$), we retained spectra covering the full green arm at 5160--5730~\AA\ and the red range from 6100 to at least 6680~\AA\ of the nominal 6100--6790~\AA\ red arm of 4MOST. Such data are available for 371 stars in the initial sample. For 82 stars, 528 spectra also cover the full three-arm 4MOST HR setup, including the blue arm at 3926--4355~\AA. The typical parameter space expected for Cepheids, including their variations due to pulsation, is $4000 \lesssim T_{\rm eff} \lesssim 7000~\mathrm{K}$ and $0.0 \lesssim \log g \lesssim 2.5~\mathrm{dex}$ \citep{Vasilyev18}, which is fully covered by our methodology. To provide a homogeneous input set for testing our code in the context of the 4MIDABLE-HR survey, we degraded all archival spectra to a common resolving power of $R\approx20\,000$ using an FFT convolution\footnote{\url{https://github.com/stormnick/one_payne/blob/main/src/convolve.py}}. Although real 4MOST spectra will have wavelength-dependent resolution, even the latest version of the 4FS ETC tool (4MOST Facility Simulator Exposure Time Calculator) uses a simple convolution to simulate the 4MOST system response. We also corrected the spectra for telluric absorption by atmospheric $\mathrm{O}_{2}$ in the range 6270--6320~\AA\ using \texttt{Telfit} \citep{Telfit}.

Our final sample of CCs with valid atmospheric parameter solutions comprises 812 spectra of 351 stars. Of these, 113 stars have at least two visits, and 23 have five or more multi-epoch observations spanning different pulsation phases. The sample includes 433 UVES spectra, 255 HARPS spectra, and 124 FEROS spectra. Among these targets, 250 are also included in the 4MIDABLE-HR input catalog. The sample also includes 33 Type II Cepheids, 23 RR Lyrae variables and 8 Anomalous Cepheids targeted by the same programs, allowing us to test the applicability and validity of the method across different variability classes. Together, these samples amount to 975 spectra of 415 variable stars.

We complemented the spectral data of CCs with photometric and astrometric data from the literature. Specifically, we used the intensity-averaged Gaia DR3 magnitudes in the G, \Gbp and \Grp bands from \citet{Ripepi2023}, obtained using the methods described in \citet{Clementini2016}. Coordinates were obtained from the Gaia catalog \citep{GaiaCollab16, Babusiaux23}. When available, pulsation periods and reference epochs were adopted from the following catalogs, in decreasing order of priority: Gaia \citep{Eyer2023, Ripepi2023}, the Optical Gravitational Lensing Experiment survey \citep[][OGLE]{OGLEIV, Pietrukowicz2021}, the Zwicky Transient Facility \citep[][ZTF]{ZTF}, and the All-Sky Automated Survey for Supernovae \citep[][ASAS-SN]{ASASSN}. As a result, we have pulsation periods, pulsation mode classifications, average magnitudes, astrometric parameters, fundamental parameters and NLTE abundances for 335 Classical Cepheids.

Fig.~\ref{fig:MWFaceon} shows the full sample of Cepheids and other variables in this study, together with the Cepheids in the 4MIDABLE-HR input catalog.

\section{Methods}

\label{sec:methods}

\subsection{Artificial Neural Network}

\label{sec:ANN}

The pipeline that we use for the astrophysical analysis of CC spectra relies on an ANN described in detail in \citet{Storm2026}. The prototype of the code was presented in \citet{Kovalev2019, PayneYST}. Here we only provide a brief summary of its capabilities. 

The NLTE ANN was trained on a large multi-dimensional grid of NLTE synthetic stellar spectra that cover the full range of the 4MOST-HR spectrograph. The grid has 20 dimensions that include the atmospheric parameters (\teffnospace{}, \loggnospace{}, metallicity [Fe/H], microturbulence \vmicnospace{}) and the individual chemical abundances ([X/Fe]\footnote{Abundance of the given element to iron compared to the Sun in decimal logarithmic units.}) of 16 elements Mg, Ca, Si, Ti, O, Mn, Ba, Eu, Y, Co, Sr, Na, Al, Ni, in NLTE, and C and Cr in LTE. To avoid biases in the training set, spectra were generated with randomly distributed parameters within the boundaries of the grid. 

\begin{figure*}
	\includegraphics[width=\textwidth]{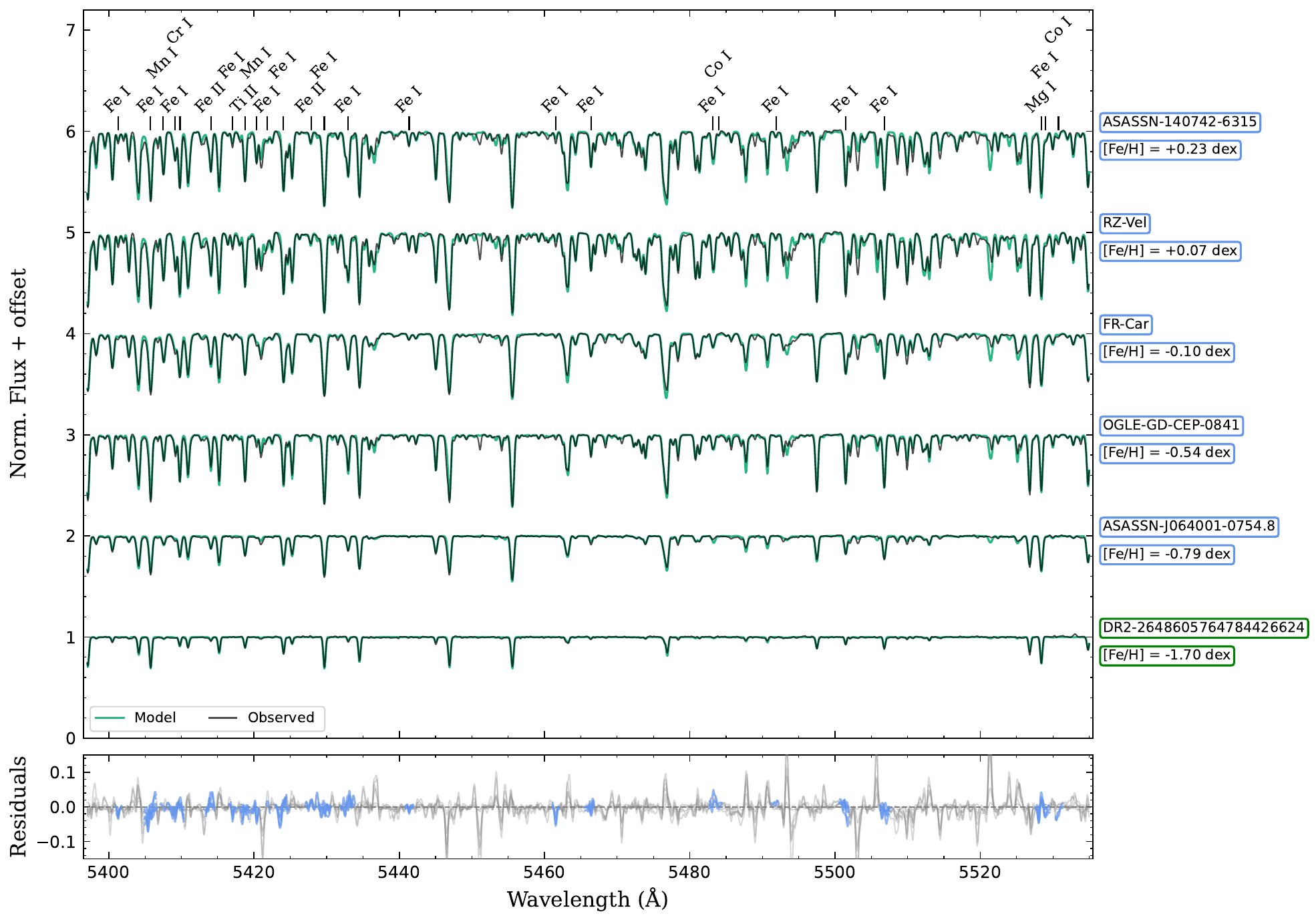}
    \caption{Selected archival spectra of classical Cepheids and an RR Lyrae star in our sample. The black and green lines show the observed and best-fit NLTE spectra, respectively. The subset of diagnostic spectral lines in this wavelength range is shown with black vertical lines. Bottom panels show the residuals for the fitted models, gray lines show the residuals across the entire range, blue segments show residuals on wavelength windows around diagnostic lines used for the best-fit optimization. Names and derived stellar metallicity are indicated on the right side.}
    \label{fig:FitShowcase}
\end{figure*}

The NLTE spectra thus cover the following parameter space following the distribution of \texttt{MARCS} stellar model atmospheres: $3500 \leq T_{\rm eff}/ \rm{K} \leq 8000$, $0.0 \leq \log{\rm g}\leq 5.0$, $-5 \leq \rm{[Fe/H]} \leq 0.5$, $0.5 \leq \rm{V}_{\rm mic} / \rm{km s}^{-1} \leq 5$. Compared to \citet{Storm2026}, we computed additionally around $45\,000$ spectra in the low \loggnospace{} regime ($0.0 \leq \log{\rm g}\leq 1.0$) up to a total of $449\,953$ spectra to better cover the relevant CCs parameter space. The NLTE spectra were computed using the TurboSpectrum \citep{Gerber2023, Storm2023} code, employing the \texttt{MARCS} 1D LTE model atmospheres \citep{Gustafsson2008}. The radial velocity ($\rm{V}_{\rm{RV}}$) and broadening (\vbrdnospace{}) are applied to the synthetic spectra during the fitting and left as free parameters of the model. At the 4MOST-HR resolving power, it is not possible to disentangle the macro-turbulence and rotational broadening. Moreover, the highly dynamic atmospheres of Cepheids show complex interaction of convection and pulsation \citep{Anderson2016c, Anderson2016d, Vasilyev18}. Therefore, we encapsulate all these sources of broadening in the single parameter \vbrdnospace{}, which was included as a free parameter of the model and approximated as a single-component Gaussian profile. The atomic and molecular line list was adopted from the Gaia-ESO line list \citep{Heiter2021, Gerber2023}. However, some updates were applied as described in \citet{Magg2022}, and VALD3 database \citep{Ryabchikova2015} was used to fill the gaps that were not covered by the Gaia-ESO line list. NLTE line formation was included using the NLTE departure coefficient grids calculated per element using the MULTI2.3 \citep{Carlsson1986} code. NLTE departure grids were computed using the NLTE atomic models developed and validated by our team and by collaborators for H \citep{Mashonkina2008}, O \citep{Bergemann2021}, Na \citep{Ezzeddine2018}, Mg \citep{Bergemann2017}, Al \citep{Ezzeddine2018}, Si \citep{Bergemann2013, Magg2022}, Ca \citep{Mashonkina2017, Semenova2020}, Ti \citep{Bergemann2011}, Mn \citep{Bergemann2019_Mn}, Fe \citep{Bergemann2012a, Semenova2020}, Co \citep{Bergemann2010, Yakovleva2020}, Ni \citep{Bergemann2021, Voronov2022}, Sr \citep{Bergemann2012b, Gerber2023}, Y \citep{Storm2023, Storm2024}, Ba \citep{Gallagher2020}, and Eu \citep{Storm2024}. The grid covers typical parameter space for Cepheids, although, at ($T_{\rm eff} \gtrsim 5000$ K) and low surface gravities ($\log g \lesssim 0.5$) the training set is less dense due to fewer available model atmospheres, therefore, results in this range should be interpreted with caution. The full parameter space covered by the synthetic spectra used in the training of the ANN emulator can be seen in Fig. \ref{fig:paramspace}.

\subsection{Chemical abundance analysis}
\label{sec:fitting}

The analysis of stellar parameters and chemical abundances of CCs is performed via three main steps. In the first step, we normalize the continuum. We note that this step is only needed in the context of this paper in application to the archival UVES, FEROS and HARPS data, as the spectral data analysis in 4MOST will be carried out using the 4MOST 4GP continuum normalization procedure. Since the 4MOST normalization routine is still in active development, we normalize spectra to the best of our ability and leave the normalization uncertainty tests to the upcoming paper by the 4MOST Galactic pipeline working group. In this work, we find the best-fit continuum spectrum model by fitting for $T_{\rm eff}$, $\log{\rm g}$, \ironab, \vmicnospace{}, and \vbrdnospace{} over broad wavelength ranges in the observed spectra. We divide the observed spectrum by this synthetic template, and set the continuum using a median filter and a first-degree polynomial fit every 5 nm, obtaining the normalized spectra. To limit residual biases, we restrict the fit to lines that have been carefully inspected over most of the parameter space of the 4MIDABLE-HR sample. We remove lines whose depths are less than 3\% of the continuum, or are not well represented by a simple Gaussian profile. This list of key diagnostic lines is included in Table \ref{tab:lines}.

In the second step, we perform the astrophysical analysis of atmospheric parameters and abundances with a $\chi^{2}$ minimization procedure, composed of distinct steps. We use the \texttt{minimize} function from Python package \texttt{LMFIT} \citep{Newville2025} to obtain the best fit parameters and covariance matrix at each step. First, we fit for \teffnospace{}, \loggnospace{}, \ironab, \vmicnospace{}, \vbrdnospace{}, Mg and Ca abundances. This analysis is carried out using the 137 lines of Fe I, Fe II, and the 5 Mg I and 11 Ca I lines detailed in Table \ref{tab:lines}, simultaneously. The Fe I and Fe II lines provide sensitivity to the atmospheric parameters through both the ionization balance and the implicit excitation balance of the more numerous Fe I lines, while adding additional constraints from Mg and Ca lines. We then fix \teffnospace{}, \loggnospace{}, \vmicnospace{} and \vbrdnospace{} and refit \ironab individually over each line mask for Fe I and Fe II lines. The final value for the metallicity is obtained from the average [Fe/H] of all the individual lines, after a 3-$\sigma$ clipping over the per-line metallicities.

Finally, we determine the individual elemental abundances using clean diagnostic lines separately for each element, that is, using the trace element assumption. This is a common approach in the literature \cite{Nissen2018, Jofre2019} and it is justified, since the model atmospheres of stars, as well as the grids of NLTE departure coefficients, are discretized only on metallicity and alpha element enhancement. Due to the potential blending at $R \sim 20{,}000$ for the higher metallicity stars, we perform one extra iteration on the abundances, one at a time and fixing the rest of the abundances to the values obtained in the previous step. We chose not to include hydrogen lines in the optimization, as these are not well described by classical 1D hydrostatic equilibrium model atmospheres for giant stars \citep{Bergemann2016, Wedemeyer2017}, and some variable stars including Cepheids can exhibit emission in H lines \citep{Hutchinson75, Gillet14}.

Additionally, we flag results for which any of the atmospheric parameters or abundances reaches the edge of the training set grid. This is especially important for long-period Cepheids (i.e. P $\gtrsim  35$ days), whose surface gravities are near or even over the edge of the training set grid of $\log{\rm g} \sim 0.$ \citep{Molinaro2011, Vasilyev18}. Depending on the clipped parameter, this procedure either flags the entire spectrum as unreliable, in the case of atmospheric parameters, or only the corresponding elemental abundance. For reliable spectra, a final \chisq{} is evaluated over the observed coverage. Uncertainties are taken from the covariance matrix in the last stage and added to the systematic errors in atmospheric parameters and abundances reported by \citet{Storm2026}. 

\begin{figure}
	\includegraphics[width=\columnwidth]{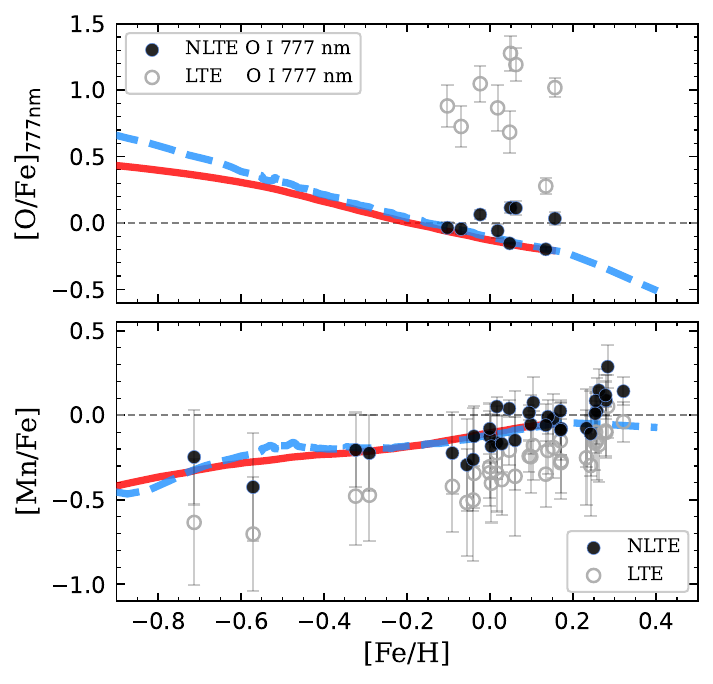}
    \caption{Abundance ratios [O/Fe] and [Mn/Fe] in NLTE and LTE as a function of metallicity for CCs in our sample. NLTE and LTE-based abundances are marked with filled black and open gray circles symbols, respectively. Error bars are the standard deviation between the abundances for individual lines. OMEGA+GCE models from \citet{Lian2023} are over plotted with star formation efficiency of $\epsilon_*$: 0.03 (solid red) and $\epsilon_*$: 1 (dashed blue). This illustrates the sensitivity of O I 777 nm lines and Mn lines to NLTE effects. }
    \label{fig:NLTE_TSFitPy}
\end{figure}

Following \citet{Storm2026}, we validate our \teffnospace{} and \loggnospace{} scale by comparing our best-fit parameters with the Gaia FGKM benchmark stars from \citet{Heiter2015b}. This provides an external check based on \teffnospace{} values derived from angular diameters and bolometric fluxes, and \loggnospace{} values inferred from stellar radii and stellar evolution models. The spectra were taken from the library compiled by \citet{Blanco-Cuaresma14}, including 65 FEROS, NARVAL, HARPS, and UVES spectra for 30 stars. They were homogenized to the 4MOST-HR resolving power and wavelength coverage using the same procedure described in Sec.\ref{sec:Data}. This benchmark sample covers the FGKM parameter space over the ranges $3600 \lesssim T_{\rm eff}/\mathrm{K} \lesssim 6800$, $0.3 \lesssim$ \logg $\lesssim 4.8$, and $-2.9 \lesssim \mathrm{[Fe/H]} \lesssim 0.3$ (see Fig. \ref{fig:APP_biases}). We analyze these spectra with the fitting method described in Sec.\ref{sec:fitting} and compare the resulting parameters with the values from \citet{Heiter2015b}. This comparison reveals systematic offsets of $-30 \pm 100$ K in \teffnospace{} and $-0.1 \pm 0.3$ dex in \loggnospace{}. Similar biases have been reported in previous NLTE Payne-based studies \citep{Gent2022SAPP, Storm2026}. We therefore apply constant bias corrections to \teffnospace{} and \loggnospace{} at the end of the first fitting step, and use the corrected parameters to determine \ironab{} and the individual elemental abundances, as described above. We show the Kiel diagram and the comparison with \citet{Heiter2015b} in more detail in appendix \ref{app:biases}.

As an additional validation step, we compare the average abundances of the target elements obtained from seven solar spectra observed with UVES, HARPS, FEROS and NARVAL with the solar abundances from \citet{Magg2022}. In Table \ref{tab:all_solar}, we show the mean abundances and standard deviations from our analysis of the solar spectra. These values are consistent with the NLTE solar abundances from \citet{Magg2022} within 0.1 dex for most elements. To normalize the values derived by the pipeline to the adopted solar abundance scale, we correct all pipeline-derived results using the differences between \citet{Magg2022} abundances and those listed in Table \ref{tab:all_solar}. This procedure is explained in detail in appendix \ref{app:biases}.

\begin{figure*}[h]
	\includegraphics[width=\textwidth]{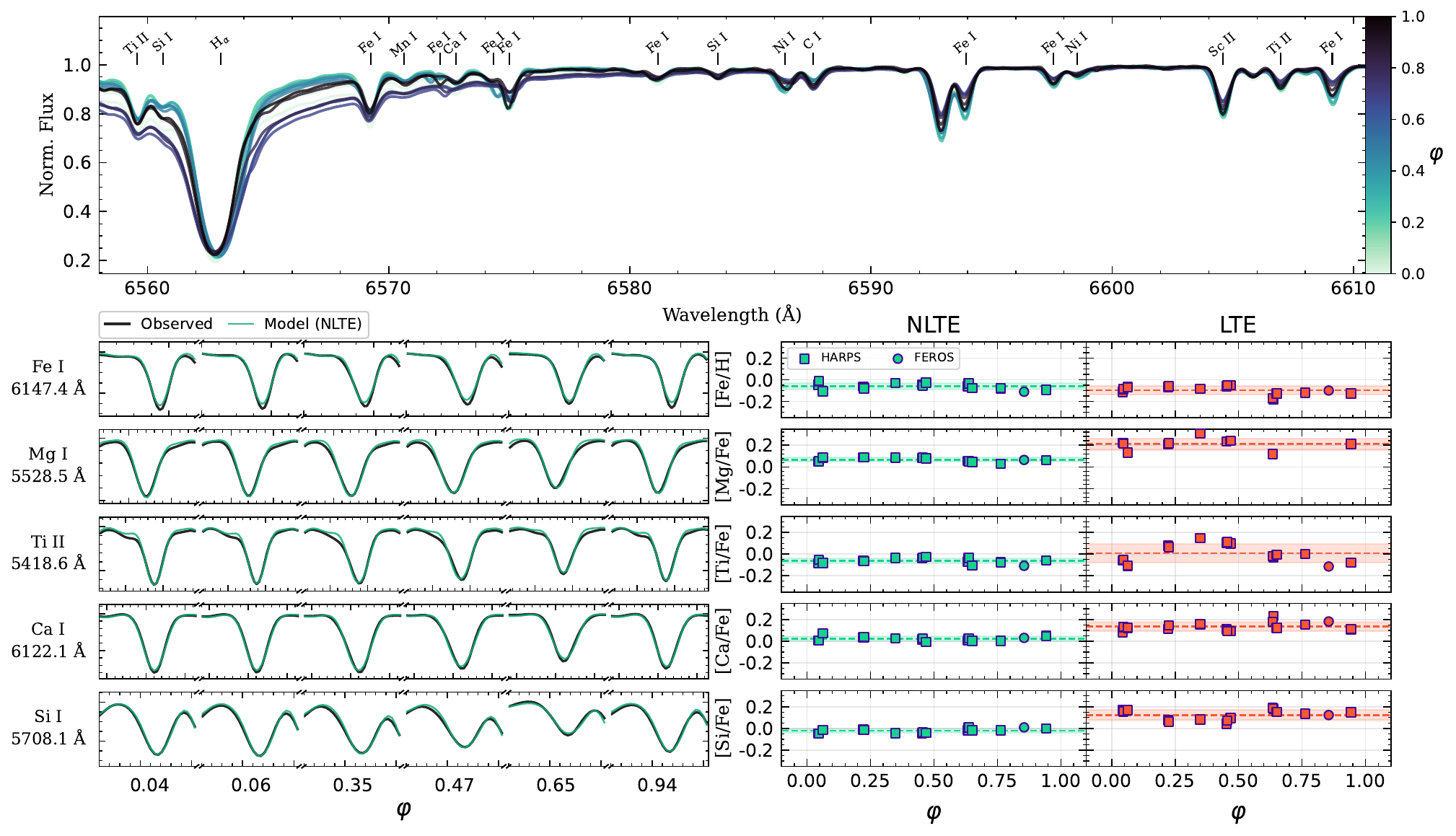}
    \caption{Multi-epoch spectra of CC R TrA in NLTE and LTE. Top: Portion of the spectra in the vicinity of the $\rm{H}_{\alpha}$ line; the colors indicate the normalized spectra at different pulsation phases. Lines present in our line list are overplotted. Bottom left: Line profiles of key diagnostic features across phase, with the observed spectra shown in black and the best-fitting model in green. Bottom right: Phase dependent abundances for a selection of elements. Light green points on the left represent NLTE abundances for the phases available, while red points on the right correspond to LTE abundances. The NLTE results demonstrate stable solutions in spite of the stellar variability. Only measurements with both stellar parameters and abundances lying within the grid limits are displayed. Dashed lines indicate phase-averaged values, and the shaded regions denote the standard deviation of abundances in NLTE (green) and LTE (red).}
    \label{fig:CalibMuliphase}
\end{figure*}

Figure~\ref{fig:FitShowcase} illustrates the procedure on Cepheids and an RR Lyrae star spanning a wide range of metallicities. The observed spectra were homogenized to the nominal 4MOST resolving power of the 4MOST-HR spectrograph and fitted with the aforementioned procedure. This showcases the applicability of the method to Cepheids and other variables across a broad range of metallicities. With this procedure, we obtain atmospheric parameters and abundances for 975 spectra in our sample corresponding to 415 individual stars, determined self-consistently in NLTE. 

To illustrate the impact of NLTE effects on the GCE interpretation of Cepheid abundances, we calculate the abundance ratios [O/Fe] and [Mn/Fe] as a function of metallicity. We use both LTE and NLTE-based abundance ratios on FEROS spectra that include the O I 777 nm triplet. The results of this comparison are shown in Fig. \ref{fig:NLTE_TSFitPy}. To isolate the influence of NLTE effects, we calculate all abundances with \texttt{TSFitPy} first in LTE and then in NLTE, using the same atmospheric parameters and departure coefficient grids as in the method described in this Section. This approach is justified as, in this metallicity range, we do not expect large differences in atmospheric parameters due to NLTE effects \citep{Kovalev2019}. We find that neglecting NLTE effects systematically shifts the O abundances by more than 0.5 dex, while in the case of Mn the shift is in the order of -0.2 dex. This is also in good agreement with the calculations from \citet{Sitnova2013} and \citet{Bergemann2008}\footnote{NLTE corrections calculated in \nolinkurl{nlte.mpia.de}} in the case of oxygen and manganese, respectively.

\subsection{Validation of atmospheric parameters and abundances}
\label{sec:Validation}

To test the consistency and accuracy of our method, we performed a series of validation tests. These are based on the multi-phase analysis of atmospheric parameters of Cepheids (\teffnospace{}, \loggnospace{}, R) and the consistency of chemical abundances as a function of phase for our 23 stars with multi-phase spectra. We test the consistency of solutions in Sect. \ref{sec:Validation_parameters}, and a comparison with previous literature studies performed in LTE and in NLTE in Sect. \ref{sec:comparison}. We also assess the accuracy of oxygen abundances in our method in Sec. \ref{sec:oxygen}.

\subsubsection{Stellar parameters and abundances}
\label{sec:Validation_parameters}

First, we analyzed the stars in the multi-phase sample. This sample of CCs comprises 23 stars with valid atmospheric parameter solutions for 5 or more spectra, covering the ranges $4100 \lesssim T_{\rm eff}/\mathrm{K} \lesssim 6600$, $0.4 \lesssim$ \logg $\lesssim 2.3$, and $-0.35 \lesssim \mathrm{[Fe/H]} \lesssim 0.20$.

Due to changes in stellar structure of CCs throughout pulsation, we expect atmospheric parameters to change during phase, affecting the departures from LTE \citep{Vasilyev2017, Vasilyev18}. In Fig. \ref{fig:CalibMuliphase}, we show an example comparison of multi-phase abundances and metallicities for the Cepheid R-TrA using LTE and NLTE spectral models. The latter values are computed as described in Sec. \ref{sec:methods}, whereas the LTE values are computed using the TSFitPy code \citep{Gerber2023, Storm2023}, fixing \teffnospace{}, \loggnospace{}, \vmicnospace{} and \vbrdnospace{} to the values derived with our method, and fitting the abundances on the same spectral windows. The bottom panels in Fig. \ref{fig:CalibMuliphase} show changes in line profiles across phase and the corresponding abundances in NLTE and LTE. From this comparison, we see that NLTE values show lower dispersions. This is expected, as such a systematic increase in dispersion caused by the larger biases in LTE-based analysis is known for different chemical elements in FGK-type stars (see e.g. Mg in \citealt{Bergemann2017}, Y in \citealt{Storm2024, Guiglion24}, Ca in \citealt{Semenova2020}). 

Identifying the origin of these deviations on a line-by-line basis is difficult, as NLTE effects are sensitive to the details of line formation, including optical depth, source function, and behavior of the departure coefficients for the lower and upper energy levels. These are fundamentally different in LTE and NLTE radiative transfer calculations, even in 1D model atmospheres. Across all 23 stars, for the elements shown in Fig. \ref{fig:CalibMuliphase}, the average abundance differences range from -0.05 dex for Si, to 0.02 dex in the case of Ti, consistent with expectations from giant stars \citep{Bergemann2011, Bergemann2013}. Comparing the standard deviation of abundances across phase, all elements show lower dispersions in NLTE, being on average 0.06 dex lower. We include more examples of this comparison for stars with more than ten spectra in App. \ref{app:MultiphaseAll}.

In Fig. \ref{fig:multiphase_5_show}, we show the phase-dependent atmospheric parameters for six multi-phase stars. For these stars, \teffnospace{} varies by 500–1000 K across the cycle, while \loggnospace{} varies by 0.3–0.7 dex, with average values consistent with CCs. To compare our results with the theoretical position of the instability strip from stellar structure models, we convert the observed radial velocities into pulsation velocities and estimate the mean radius of each star. We use the projection factor and period–radius relation from \citet{Trahin2021} to calculate the stellar radius as a function of phase. We then fit Fourier series to the phase-dependent stellar radius, $R$, and \teffnospace{}, and use the Stefan–Boltzmann law to estimate the luminosity, $L$. The resulting distribution of our sample in the \teffnospace{}–$L$ plane is compared with the instability strip predictions for solar metallicity from \citet{Anderson16} in the bottom panel of Fig. \ref{fig:multiphase_5_show}.

We find good agreement between the pipeline results and the models for the instability strip, with the values and relative changes in \teffnospace{} and \loggnospace{} consistent with expectations from stellar pulsation. We note that the excursions beyond the instability strip are not unexpected, as Cepheids may approach or temporarily cross the hot edge of the instability strip during pulsation, while mostly remaining within the cool boundary \citep[see Fig.~3 of][]{Javanmardi2021}. The cool boundary is set by the increasing efficiency of convection at lower effective temperatures, which damps the pulsation by reducing the driving from the $\kappa$-mechanism in the partial ionization zones  \citep{Bono99, Anderson16}.

For the 23 stars in the sample, we find fractional radius variations of 2–18\% across the pulsation cycle, corresponding to absolute changes of $\Delta \rm{R} \approx 0.8$ to 27 \Rsun. These values are consistent with 1D pulsation models at the solar metallicity \citep{Bersier1997, Bono2000}. We also note that our results reproduce the typical behavior of CCs in the color–magnitude diagram \citep{GaiaCollab19, Anderson2025}.  

\subsubsection{Comparison with the literature}
\label{sec:comparison}

As a second validation test, we compare our results with two recent Cepheid abundance studies: \citet{DaSilva22}, based on LTE, and \citet{Nunnari25}, based on NLTE analysis. Together, these works provide up-to-date LTE and NLTE references for atmospheric parameters and chemical abundances. Using the common spectra analyzed in each study, we calculate the mean differences and corresponding standard deviations for \teffnospace{}, \loggnospace{}, and [Fe/H]. We find offsets of $\Delta$\teffnospace{} $= 72 \pm 104$ K, $\Delta$\loggnospace{} $= -0.03 \pm 0.27$ dex, and $\Delta$[Fe/H] $= -0.01 \pm 0.07$ dex among the 174 common spectra with \citet{DaSilva22}. For the 417 common spectra with \citet{Nunnari25}, we find offsets of $\Delta$\teffnospace{} $= 63 \pm 162$ K, $\Delta$\loggnospace{} $= 0.14 \pm 0.34$ dex, and $\Delta$[Fe/H] $= 0.04 \pm 0.12$ dex. Overall, there are no differences beyond the $1\sigma$ level in the reported parameters.

\begin{figure}
	\includegraphics[width=\columnwidth]{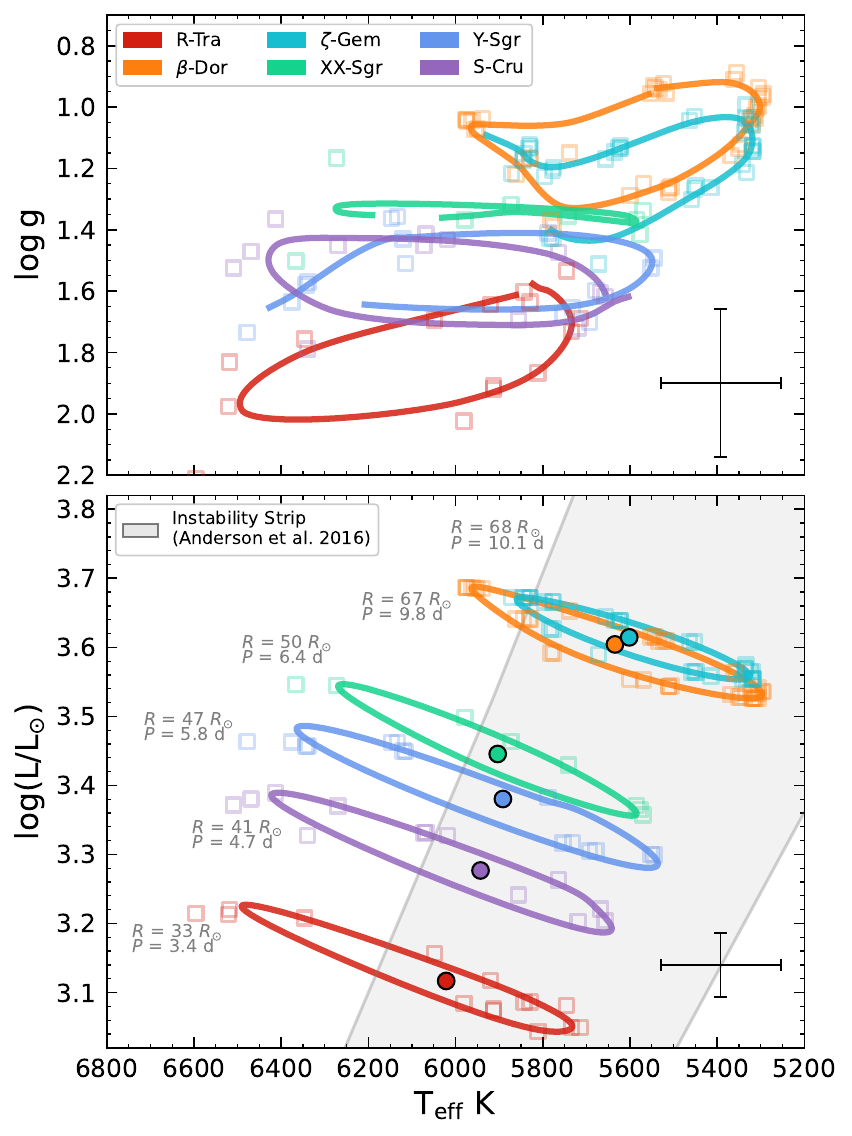}
    \caption{\teffnospace{}-\loggnospace{} diagram (top) and \teffnospace{}-luminosity diagram (bottom) of selected CCs with multiphase observations. Squares show the parameters derived from individual spectra, solid lines represent the Fourier fits to the phase progression of \teffnospace{}, \loggnospace{}, and luminosity. The gray region shows the instability strip from \citet{Anderson16} for a solar-metallicity star with a rotation rate equal to half the critical angular velocity, as defined by \citet{Georgy2013}. Filled circles show the average positions of the stars in the HR diagram across phase. Crosses show the typical uncertainties in \teffnospace{}, \loggnospace{}, and luminosity.}
    \label{fig:multiphase_5_show}
\end{figure}

\begin{figure}
	\includegraphics[width=\columnwidth]{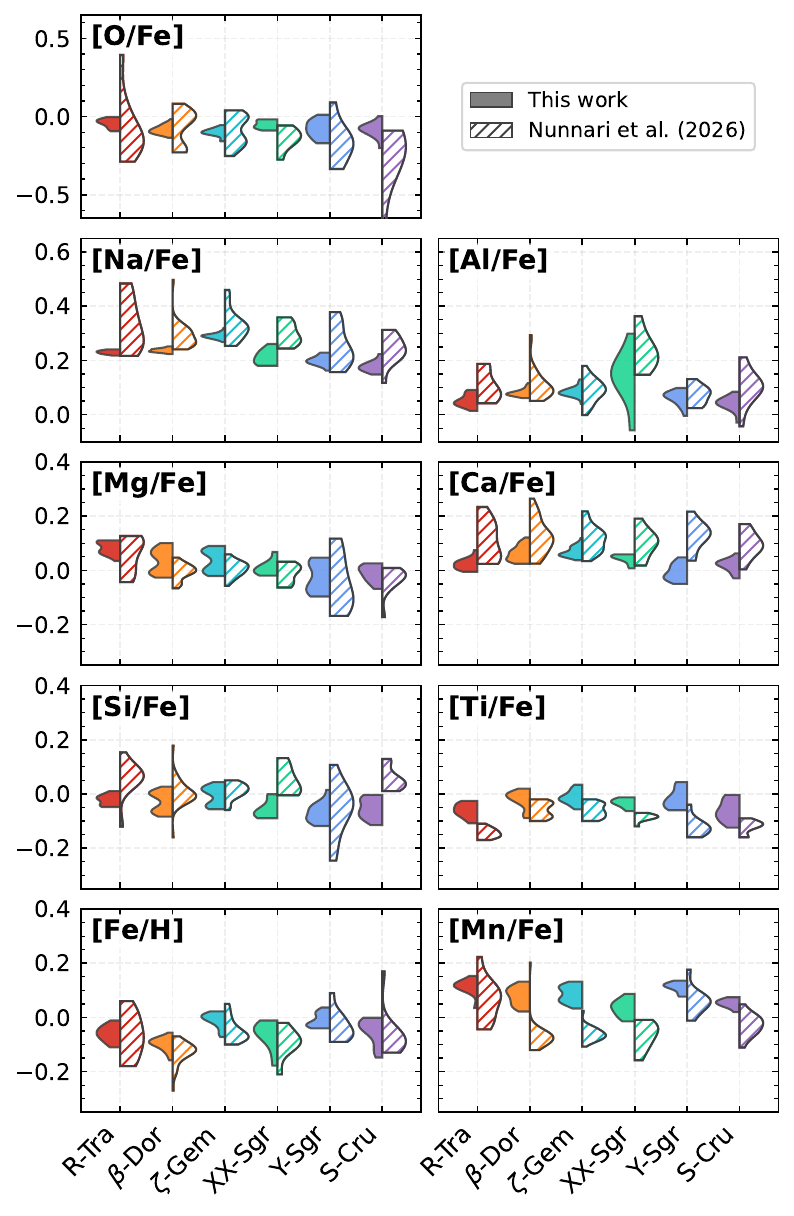}
    \caption{Abundance histograms for the 6 CCs shown in Fig.~\ref{fig:multiphase_5_show}, obtained in NLTE in this work (solid) and reported by \citet{Nunnari25} (hatched) for the spectra common to both studies. Each panel shows the NLTE abundance distribution of an individual element, shown in the top left.}
    \label{fig:multiphase_5_abundances}
\end{figure}

To compare our derived NLTE abundances with independent NLTE results, we show in Fig. \ref{fig:multiphase_5_abundances} the NLTE abundance distributions for the elements in common with \citet{Nunnari25}, namely O, Na, Mg, Al, Si, Ca, Ti, Mn, Fe. The comparison is restricted to the same stars used in Fig. \ref{fig:multiphase_5_show}. Despite differences between the two analyses, including the resolution, adopted line lists, atmospheric-parameter determination, spectral coverage, and NLTE model atoms, the abundance distributions are broadly consistent.

We further assess the internal consistency of our NLTE abundances by quantifying how much they vary across pulsation phase. For each of the 23 multiphase stars, we compute the standard deviation across phase of the abundance ratios presented in Fig. \ref{fig:multiphase_5_abundances}. Taking the average of the phase to phase dispersions for those elements gives a mean value of $\sim 0.057\,\mathrm{dex}$. Restricting the comparison to matched spectra only, and averaging over the common target elements, gives mean dispersions of approximately $0.042$ dex for our results, and $0.063$ dex for \citet{Nunnari25}. On these matched observations our values show smaller mean dispersions, despite our narrower wavelength coverage and lower spectral resolution. For $[\mathrm{Fe/H}]$ alone, our mean dispersions are lower than those of \citet{DaSilva22} and \citet{Nunnari25} by $0.005$ and $0.028\,\mathrm{dex}$, respectively, using the spectra shared with each study.

\begin{figure}
	\includegraphics[width=\columnwidth]{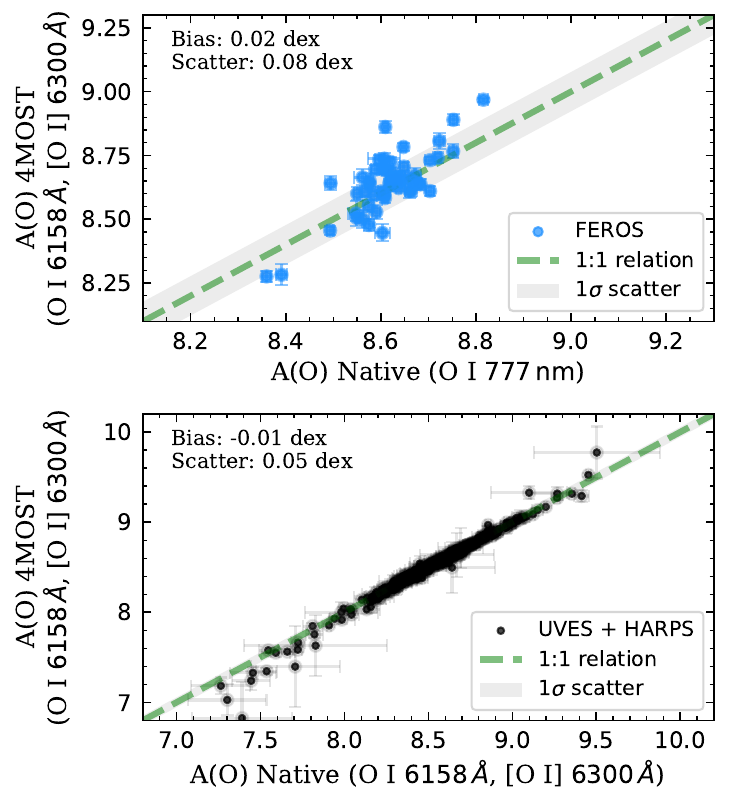}
    \caption{Comparison of NLTE oxygen abundance for stars in our sample in both native resolution and 4MOST resolution with different lines. Top panel shows the comparison between O abundance derived from the 777 nm O I lines at the native FEROS resolution and the derivation from O lines in the 4MOST range and resolution. Bottom panel shows a comparison for UVES and HARPS spectra in our sample at their respective native resolution, using the [O I] line at $\sim 6300 \, \text{\AA}$ and the $\sim 6158 \, \text{\AA}$ lines in both cases. Green dashed line shows a 1:1 relation and shaded region shows the 1$\sigma$ scatter for both panels.}
    \label{fig:Oxygen}
\end{figure}

We note that the consistency of abundance results as a function of phase is important as timing the 4MOST observations to specific pulsation phases is not feasible, thus, only spectra taken at random phases will be available. As demonstrated with our approach, we consistently find phase independent (see bottom right panels of Fig. \ref{fig:CalibMuliphase}) solutions with typical phase dispersion of $\lesssim 0.1$ dex for Mg, Ti, Ca and Fe. The average dispersion for all the target elements for the multi-phase sample of stars can be found in Table \ref{tab:abundance_dispersion}.

\subsubsection{Oxygen abundance}
\label{sec:oxygen}

Among the chemical elements, O is especially relevant as this is the preferred tracer of chemical enrichment for Cepheids in distant galaxies \citep{Riess2021, Riess2022}. The only indicators of O in the wavelength range of the 4MOST-HR data are the 6300 \AA \, forbidden [O I] line and a much more blended O I absorption feature at 6158 \AA \, \citep[see][]{Kordopatis23}. These lines are extremely weak in spectra of FGK-type stars \citep{Caffau2015, Amarsi2016, Bergemann2021}, also their wavelength range is contaminated by telluric absorption. On the other hand, these O I lines are much less sensitive to NLTE effects than the 777 nm triplet \citep{Kiselman2001, Caffau2008, BertrandeLis2015, Bergemann2021}, and have already been used for O abundance determination of CCs for this reason \citep{DaSilva23}.

\begin{figure*}
	\includegraphics[width=\textwidth]{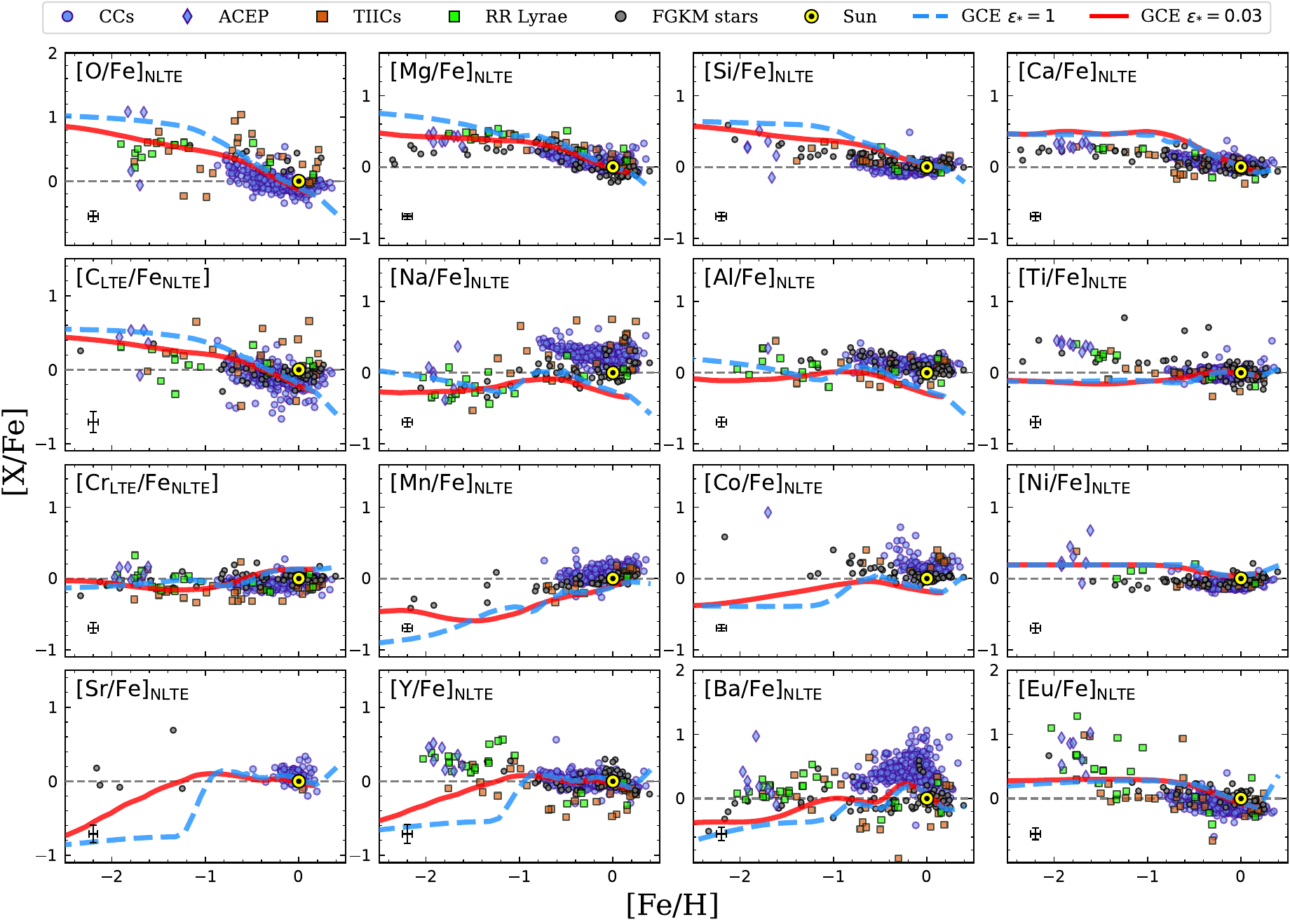}
    \caption{Chemical abundance ratios for Classical Cepheids and other variable stars. Each panel shows the abundance distribution relative to iron for a given element in NLTE, with the exception of C and Cr, which are determined in LTE. We show Classical Cepheids (blue circles), Anomalous Cepheids (blue diamonds), RR Lyrae stars (green squares), and Type II Cepheids (orange squares) from the archival sample. OMEGA+ Galactic chemical-evolution models from \citet{Lian2023} are over plotted for star formation efficiencies of $\epsilon_* = 0.03$ (solid red line) and $\epsilon_* = 1$ (dashed blue line). Dark gray points show the NLTE abundances of FGKM stars from \citet{Storm2026}. The yellow point marks the solar abundance at $(0,0)$ dex, and the black error bar indicates the average uncertainty of the measurements.}
    \label{fig:distributions_abundances}
\end{figure*}

To benchmark our oxygen abundance measurements, we perform the same analysis on the native FEROS spectra, which include the 777 nm O~I triplet lines. In contrast to our standard approach, here we fit the complete FEROS spectra at $R\sim 48\,000$ using an equivalent NLTE Payne emulator of stellar spectra, trained at $R=80\,000$. This fit includes the wavelength region around the 777 nm lines, and we adopt the same atmospheric parameters for the same spectra in all comparisons. In the top panel of Fig. \ref{fig:Oxygen}, we compare abundances obtained at the standard 4MOST-HR resolution (y-axis) with those inferred from the 777 nm O I triplet measured at the native FEROS resolution (x-axis). To also test for artifacts due to the different resolutions, in Fig. \ref{fig:Oxygen} we show the comparison using all HARPS and UVES spectra, which do not cover the 777 nm triplet, using the 6158 \AA\, O I and 6300 \AA\, [O I] lines only at their respective native resolving powers. We keep only spectra with robust atmospheric parameters and with no clear telluric contamination near the 6300~\AA\ line or issues in continuum placement. This leaves 94 FEROS spectra corresponding to 54 Cepheids.

The FEROS comparison in the top panel of Fig. \ref{fig:Oxygen} shows that A(O) derived from the 777 nm triplet at the native FEROS resolution agrees with the values obtained at 4MOST resolution from the 6158 \AA\, O I and 6300 \AA\, [O I] lines with a scatter of 0.08 dex. We emphasize that most objects are located close to the Sun’s Galactocentric radius, and therefore mainly sample near-solar metallicities and oxygen abundances close to our solar reference from \citet{Magg2022}. The same line comparison between the different resolving powers yields no significant offset. Deviations are mainly apparent at the edges of the distribution, where the higher resolving power better separates the oxygen features and helps with continuum determination.

We conclude that our method, based on the O lines in the 4MOST range, is consistent with the abundances derived from the well studied O I 777 nm lines, providing phase independent abundances across phase (see top panel of Fig. \ref{fig:multiphase_5_abundances}), which is crucial for interpreting the 4MOST data.

\section{Results}
\label{sec:results}

Here we present the results for our NLTE chemical abundance ratios, obtained using the described method and the archival high-resolution spectra of CCs, Anomalous Cepheids, RR Lyrae and type II Cepheids data. For stars observed more than once, we compute each abundance as an uncertainty weighted mean over all available spectra with reliable atmospheric parameter solutions. The full table containing the results for all individual spectra can be found in Table \ref{tab:Catalog}, and the catalog of the combined observations is shown in Table \ref{tab:AveragedCatalog}.  

\subsection{Abundance trends}
\label{sec:trends}

We present the NLTE [X/Fe] abundance ratios as a function of [Fe/H] for CCs and other variable stars in our sample in Fig. \ref{fig:distributions_abundances}. We compare the resulting trends with predictions from the OMEGA+ GCE model \citep{Cote2017, Trueman2022}, computed for different values of the star formation efficiency, $\epsilon_{\star}$. We adopt the standard value $\epsilon_{\star} = 0.03$ \citep{Lian2023} and also consider $\epsilon_{\star} = 1$ for comparison, while keeping all other parameters at their default values, according to our previous work in \citet{Lian2023}. Following \citet{Hegedus2025} and \citet{Storm2026}, we shift the theoretical [Mg/Fe] ratios from the GCE model upward by $0.3$ dex, to match the misfit at low metallicity. We also include the NLTE abundance ratios reported by \citet{Storm2026} for 84 FGKM-type stars, compiled from the careful analysis from \citet{Heiter2015b} together with nearby bright stars from \citet{Fuhrmann1993, Fuhrmann1998, Gehren2004, Gehren2006, Bergemann2008}. As shown in Fig. \ref{fig:distributions_abundances}, our NLTE abundance distributions for CCs, Anomalous Cepheids, Cepheids II, and RR Lyr are, overall, in broad agreement with the GCE models and with NLTE abundances from FGKM stars, with expected deviations due to stellar evolution effects and from the distinct stellar populations examined.

\begin{figure*}
	\includegraphics[width=\textwidth]{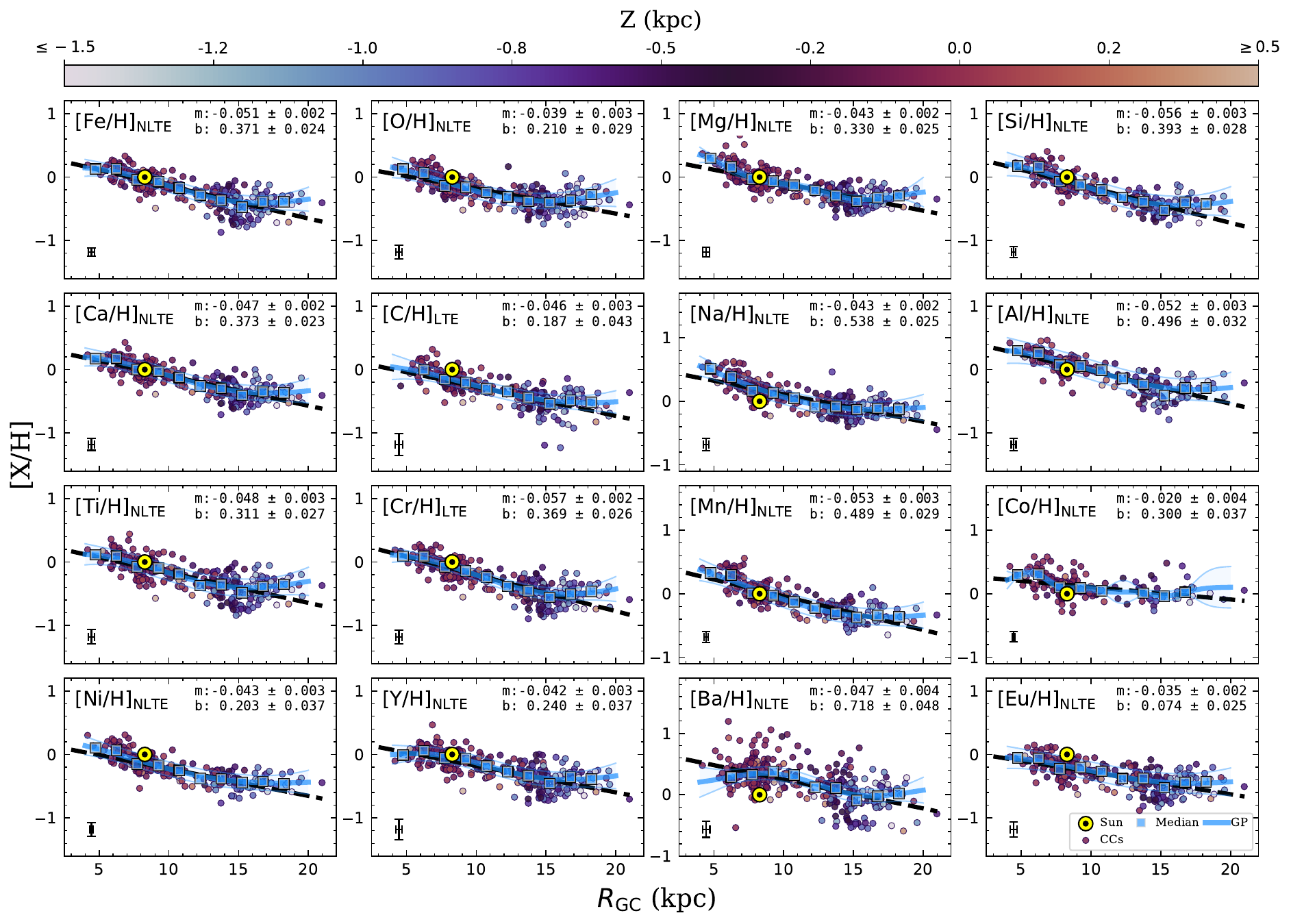}
    \caption{Radial abundance gradients for 16 elements in our sample. Abundance ratios relative to H for all elements calculated self-consistently in NLTE, except for C and Cr, which are determined in LTE. Classical Cepheids are shown as dots color coded by their height from the galactic plane. Blue squares show the running median for these stars for each element shown in the top right of each panel. Black dashed line shows the linear fit performed on the data. Blue lines show the Gaussian Process fit on the running median, with the corresponding $3\sigma$ confidence intervals as blue shaded regions. Average errors in both galactocentric distance $\mathrm{R}_{\rm GC}$ and respective abundance are shown as error bars in the bottom left. The coefficients of the linear regression of the form [X/H] = m $R_{\rm{GC}}$ + b are shown in the top right panels, together with the standard deviation of the bootstrap distributions for their uncertainties.}
    \label{fig:Gradients}
\end{figure*}

For [C/Fe], we find a significant scatter, likely related to stellar evolutionary effects depleting surface carbon in evolved stars \citep{Takeda2013}, together with observational uncertainties including continuum placement and line-mask determination. C abundances in CCs in our sample are based on the $\rm{C}_{2}$ molecular feature at 5633 \AA\ and on C I atomic lines at 6586 \AA\ and 6670 \AA, which are more sensitive to the continuum determination. The $\alpha$-elements O, Mg, Si, Ca, and the $\alpha$-like element Ti, show the increasing trends for lower metallicity in both CCs and the other variable star types, expected for Galactic stellar populations \citep[e.g.][]{Fuhrmann1998, Gehren2004, Gehren2006, Ruchti2011, Bergemann2011, Bensby2014, Kordopatis2015, Bergemann2017}. [O/Fe] shows larger scatter than [Si/Fe] and [Ca/Fe], which are known to show very tight trends with small dispersion at any metallicity \citep{Ruchti2011, Zhao2016, storm2025}. We attribute this to the use of [O I] lines for abundance determination as discussed in Sect. \ref{sec:oxygen}. 

The proton-capture elements Na and Al qualitatively follow the GCE trend for [Fe/H] $\gtrsim -1$ dex, with significant offsets of $\sim 0.4$ dex and $\sim 0.3$ dex respectively, relative to model predictions from \citet{Lian2023}. In the case of Al, this offset is well-known from NLTE studies of FGKM-type stars, and seen in independent works \citep[e.g.][]{Gehren2004, Gehren2006, Zhao2016}. For Na however, abundances appear further enhanced by $\sim 0.2$ dex relative to the FGKM sample from \citet{Storm2026}, and also show an increase at super-solar metallicities, as seen in other NLTE studies \citep{Shi2004, Zhao2016, Amarsi2020}. Similarly to C, surface Na abundance is also affected by dredge-up products in giant stars \citep{daSilva2015}, which are expected to enhance the Na abundances by $\sim 0.2$ dex \citep{Takeda2013}. The origin of the upward trend of [Na/Fe] at super-solar metallicities remains unclear \citep{Owusu2026}.

The abundance ratios of iron-peak elements Cr, Mn, Co and Ni relative to Fe are broadly consistent with the GCE model predictions, with offsets in agreement with other studies \citep[see]{Battistini2015, kobayashi2020a, Kobayashi2020b}. In our results, Mn shows a relatively large scatter, likely due to the lines used, which either fall in the 4MOST blue arm or are blended with Fe lines in the parameter space covered by CCs (See Fig. \ref{fig:FitShowcase}). The abundance ratios for [Mn/Fe] and [Ni/Fe] are in good agreement with the NLTE patterns known from the analysis of FGK-type Galactic stars \citep{Eitner2020, Eitner2023}. Nevertheless, the NLTE abundance ratios derived in this study follow the trends reported in earlier NLTE studies \citep{Bergemann2008, BergemannCescutti2010, Bergemann2010, Battistini2016, Bergemann2019_Mn}.

\begin{figure}
	\includegraphics[width=\columnwidth]{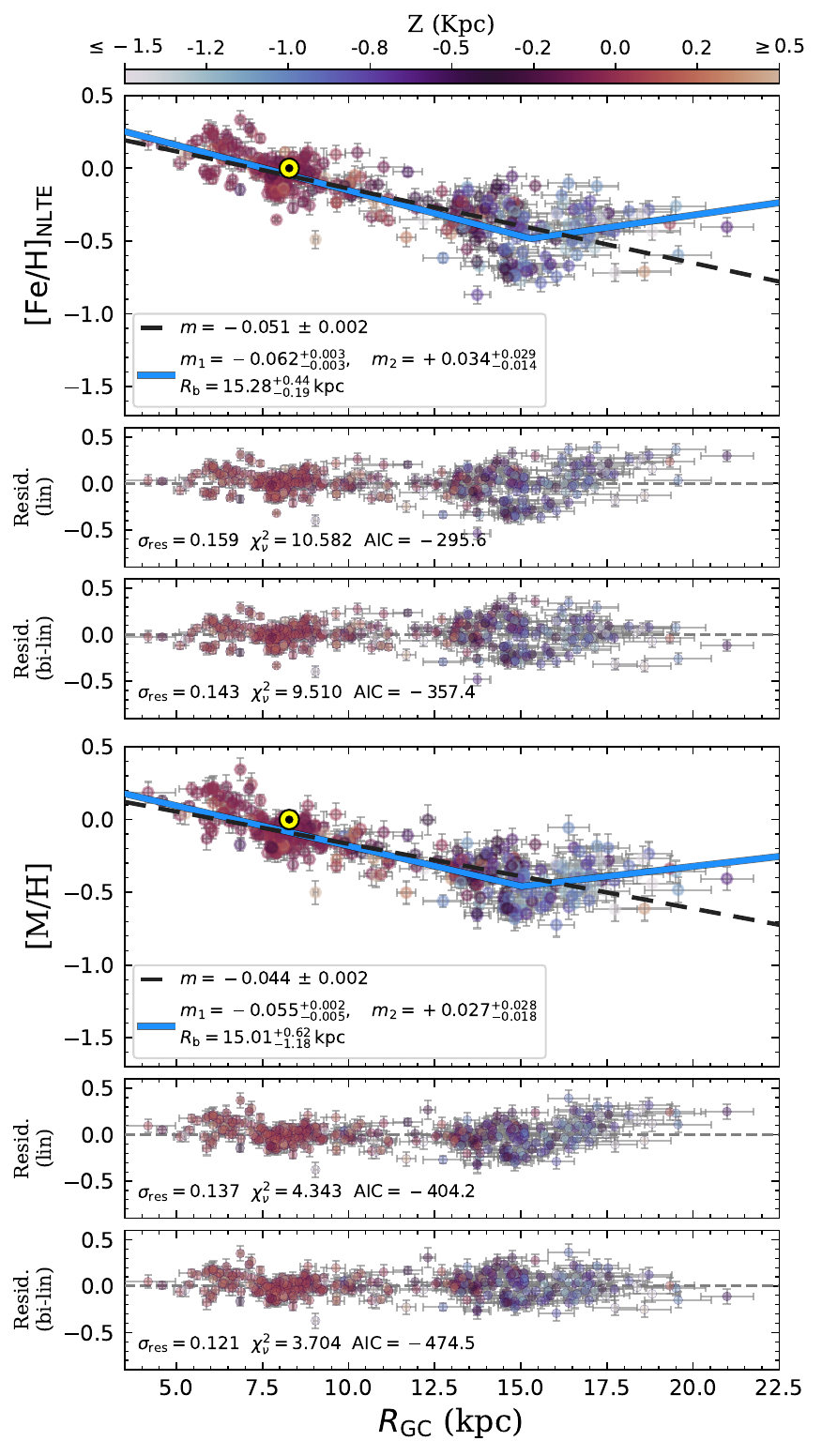}
    \caption{Linear and broken-linear fits to the Galactocentric gradients of CCs in Metallicity [Fe/H] (Top), and total metal content [M/H] (bottom). Black dashed line shows the linear fit, while the blue lines and cross show the broken linear fit and the break point, respectively. Residuals for each fit are shown in the corresponding panels, indicating the standard deviation of the residuals ($\sigma_{\rm{res}}$), the reduced $\chi^{2}$, and the Akaike Information Criterion (AIC). Yellow dot shows the solar abundance at the Galactocentric radius of 8.277 kpc.}
    \label{fig:LinBilinGrads}
\end{figure}

Neutron-capture elements produced by the s-process (Sr, Y, and Ba) and the r-process (Eu) can also be measured from the spectra of Cepheids. Perhaps the most striking feature is the enhanced Ba abundance beginning around $[\rm{Fe/H}] \sim -0.6$ dex, which is also observed in the GCE model. This characteristic ``bump'' feature has been studied before \citep{Prantzos2018, Lian2023} and signifies the transition between the dominant production channel of the s-process. The rising part of this feature is related to the change of dominant production channels, from core-collapse supernovae to the stronger enrichment by Asymptotic giant branch (AGB) stars, with a later decrease due to the metallicity dependent yields from AGB stars \citep{Bisterzo2014, Lian2023, Lugaro2023}. Eu and Y appear consistent with model expectations at [Fe/H] $\gtrsim -1$ dex. Sr abundances cluster around the solar value with some scatter, likely reflecting the use of a single diagnostic line in a region heavily blended at solar metallicity. Among the other variable stars, we find tentative evidence for low Ba abundances in metal-rich RR Lyrae stars, consistent with \citet{DOrazi26}. At $[\mathrm{Fe}/\mathrm{H}] \lesssim -1$ dex, the abundance ratios of RR Lyrae stars and Type II Cepheids appear to deviate from the GCE predictions of \citet{Lian2023}. These discrepancies may reflect differences between the populations traced by these stars and the disk populations represented by the model. Anomalous Cepheids show similar trends in $\alpha$ and iron-peak elements as those reported by \citet{RipepiACEPs}, with the main difference being that we find on average higher $[\mathrm{Y}/\mathrm{Fe}]$ values for these stars.

\subsection{Galactocentric abundance gradients}

In this section, we explore how the NLTE abundances of CCs vary with Galactocentric distance. To determine the stellar distances, we use the intensity-averaged magnitudes from \citet{Ripepi2023} to calculate the Wesenheit magnitudes in the Gaia bands as $W_{\mathrm{G}} = G - 1.90\,(G_{\mathrm{BP}} - G_{\mathrm{RP}})$. We then use the period-Wesenheit-metallicity (PWZ) relation from \citet{Ripepi2026},

\begin{equation}
    \rm{W}_{\rm{G}} = - 6.047 - 3.309 \cdot( \log{\rm{P}} - 1) - 0.527 \cdot [\rm{Fe}/\rm{H}],
\end{equation}

together with the observed Wesenheit magnitudes, the fundamentalized periods \citep{Pilecki2024}, and our NLTE metallicities, to compute the distance to each star. We estimate individual distance uncertainties by propagating the uncertainties in the photometry, NLTE metallicities, and PWZ coefficients reported by \citet{Ripepi2026}. A forthcoming paper in this series will present a detailed derivation of the PLZ and PWZ relations based on the NLTE metallicities and abundances obtained in this work. We adopt a Galactocentric distance of $8.277$ kpc for the Sun \citep{GCdist} and fit linear abundance gradients for the CC sample using maximum-likelihood estimation, accounting for the individual distance and abundance uncertainties (see Appendix \ref{app:gradients} for details).

For each element, we restrict the gradient fit to stars with a minimum number of well-defined, unblended spectral lines. Specifically, we require at least two lines for Mg, Mn, O, Na, Al, Cr, C, Y, and Ba; three for Ti, Si and Co; and five for Ca and Ni. For Eu, we adopt a one-line threshold, as the abundances for most stars are based on a single line. As a result, elements with blended or shallower lines such as Mn have comparatively fewer stars. We exclude Sr from the reported gradients because nearly all stars whose spectra cover the Sr features near $4000 \,\text{\AA}$ in the 4MOST wavelength range have Galactocentric radii within approximately $3 \,\mathrm{kpc}$ of the adopted solar value. This limited spatial coverage, combined with the uncertainties in the Sr abundance ratios, makes a meaningful gradient determination challenging. The details on the calculation of each gradient are included in the Appendix \ref{app:gradients}. The full set of linear and bi-linear gradients, together with their uncertainties and the total number of stars, is provided in Table \ref{tab:gradients}.  

Figure \ref{fig:Gradients} shows abundance gradients of 16 chemical elements: O, Na, Mg, Al, Si, Ca, Ti, Mn, Co, Ni, Fe, Y, Ba, Eu in NLTE, and C and Cr in LTE, for a sample of 335 CCs with accurate spectroscopic parameters, abundance measurements, and photometric data (see Sec.\ref{sec:Data}). The sample spans a range from $R_{\mathrm GC} \sim 4\,\mathrm{kpc}$ to $R_{\mathrm GC} \sim 21\,\mathrm{kpc}$, allowing us to trace radial abundance trends in $3$ Galactic quadrants. Points are color-coded by their vertical distance from the Galactic plane, and a slight bending at large $R_{\mathrm GC} \gtrsim 15$ kpc is visible, which is likely associated with the Galactic warp \citep{Cheng2020, Nunnari25}. Our CC sample spans the azimuthal range $\theta \in [-71^{\circ}, 95^{\circ}]$ (see Fig.~\ref{fig:MWFaceon}), with most stars located in the first quadrant, associated with the near-side warp of the disk \citep[see Fig.~4 of][]{Skowron2019}. This feature has also been observed on the far side of the disk using CCs and long-period variables \citep{Minniti2021, Albarracin2025}. 

\begin{figure*}
	\includegraphics[width=\textwidth]{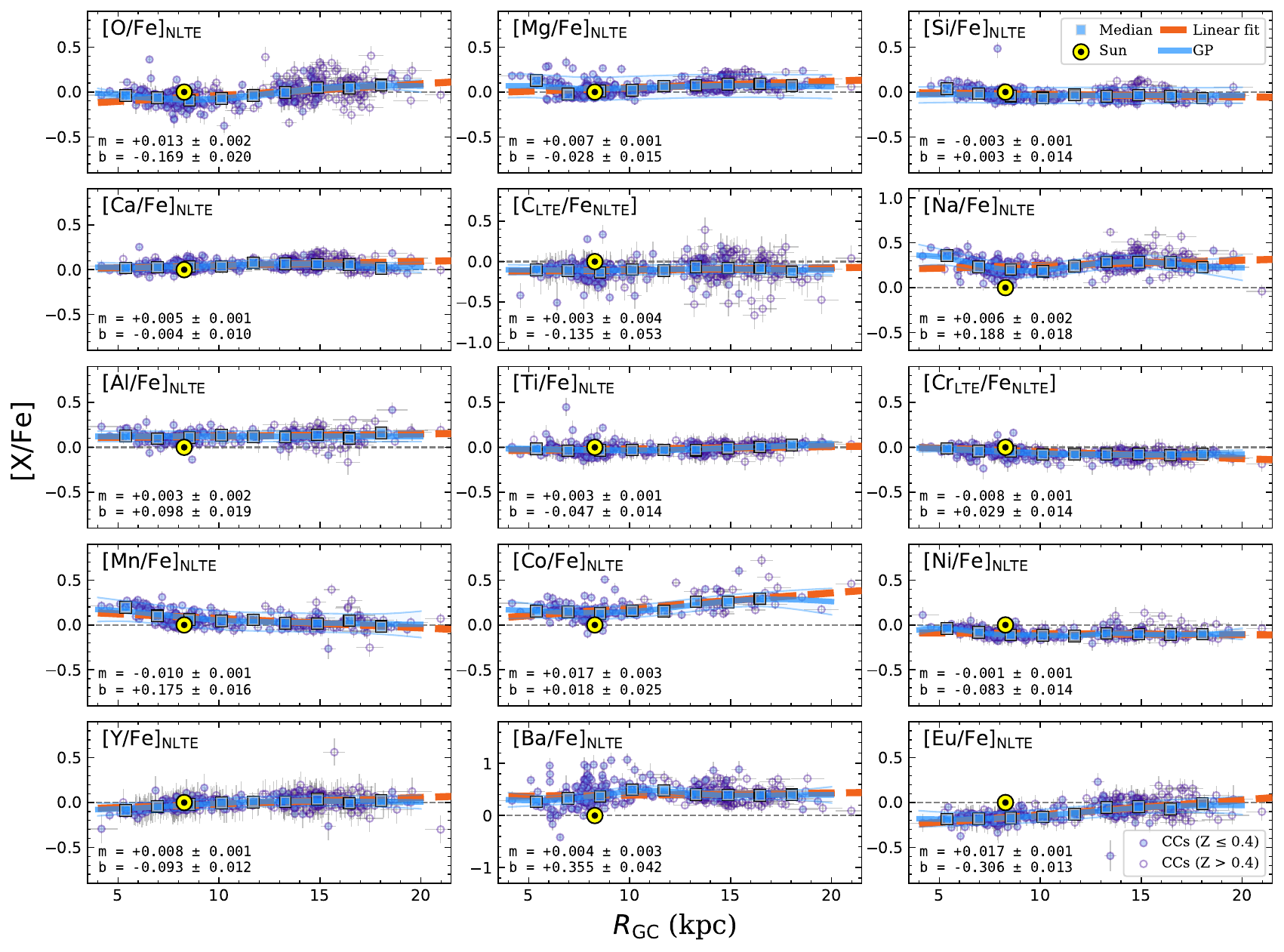}
    \caption{Radial gradients of abundances relative to iron for the target elements in our sample. All ratios are calculated in self-consistent NLTE except for C and Cr, which are determined in LTE. Blue points show elemental abundance relative to iron as a function of Galactocentric radius. Orange lines represent the linear fit to the data. Blue squares show the running median and the GP fit. Yellow dot shows the solar abundance at the Galactocentric radius of 8.277 kpc. We show the results of the linear fit of the form $[\rm{X}/\rm{Fe}] = m \, R_{\rm{GC}} + b$, together with the standard deviation of the bootstrap distributions for their uncertainties.}
    \label{fig:distributions_abundances_to_iron}
\end{figure*}

We find that the elemental abundances decrease with increasing Galactocentric radius, consistent with previous measurements of radial abundance gradients in the MW \citep{Andrievsky2002, Cescutti07, Lemasle13, Genovali14, Genovali15, Bergemann2014, Luck18, Kovtyukh2022, Trentin24b, Nunnari25} and the observed negative radial metallicity gradients in disk galaxies \citep{Ho15}. This trend is expected, since the gas density, and therefore the star formation rate, declines outward \citep{Kenicutt98}, leading to less efficient heavy-element production and weaker chemical enrichment in the outer disk \citep{Colavitti2009, Grisoni2018}. Beyond $R_{\mathrm{GC}} \sim 15\,\mathrm{kpc}$, abundance gradients tend to flatten and remain approximately constant. Similar flattening has been reported in the literature on young stellar populations, such as open clusters \citep{Carraro2007, Yong2012} and Cepheids \citep{Lemasle13, Genovali14, DaSilva23, Nunnari25}.

To characterize the flattening of the Galactic gradients, we calculate a broken-linear fit on the metallicity [Fe/H] and the total metal content [M/H], which we calculate as

\begin{equation}
[\mathrm{M/H}]
=
[\mathrm{Fe/H}]
+
\log_{10}
\left(
\sum_{i} w_{i,\odot}\,
10^{[i/\mathrm{Fe}]}
\right),
\end{equation}

where $w_{i,\odot}$ is the mass fraction contributed by element $i$ relative to the solar metal mass, and [$i$/Fe] is the abundance of $i$ relative to Fe. We compute [M/H] from the abundances of our target elements derived in Sect. \ref{sec:methods}, adopting $[\mathrm{X}/\mathrm{Fe}] = 0$ for all other metals. Figure \ref{fig:LinBilinGrads} shows both the single linear and the broken-linear fits to [Fe/H] and [M/H]. In each case, the broken linear model identifies a break in the gradient at $\rm{R}_{\rm{GC}} \sim 15\, \rm{kpc}$ and produces a smaller residual scatter, $\sigma_{\rm{res}}$, than the purely linear fit. The broken linear fits for the gradients of individual elements are listed in Table \ref{tab:gradients}. We emphasize that these fitted parameters should be treated cautiously as extra degrees of freedom, smaller sample sizes and limited radial coverage for certain elements, can make the break radius and outer slopes weakly constrained, as indicated by broad or heavy tailed bootstrap distributions. We discuss this further in App. \ref{app:gradients}.

From the simple linear fit, we obtain an [Fe/H] gradient of $-0.051$ dex kpc$^{-1}$. This value is in excellent agreement with the CC-based gradients of $-0.051$ and $-0.052$ dex kpc$^{-1}$ reported by \citet{Luck18} and \citet{Lemasle2008}, respectively. By contrast, other CC studies find either flatter or steeper slopes: \citet{DaSilva23} report $-0.041$ dex kpc$^{-1}$, while \citet{Trentin24b} derive $-0.064$ dex kpc$^{-1}$. Similarly, in a self-consistent NLTE analysis of CCs, \citet{Nunnari25} measure $-0.064$ dex kpc$^{-1}$, though their sample selection, distance scale, and methodology differ from ours.

For open clusters with ages comparable to CCs, the reported gradients cover a broad interval, ranging from $-0.038$ dex kpc$^{-1}$ \citep{Magrini23} to $-0.098$ dex kpc$^{-1}$ \citep{Otto2026}. For FGK type stars of comparable ages and Galactic-plane height within 3 kpc of the Sun, \citet{Bergemann2014} found a gradient of $-0.068$ dex kpc$^{-1}$. This is close to our inner disk [Fe/H] gradient of $-0.062$ dex kpc$^{-1}$ obtained from the broken-linear fit, shown in Fig. \ref{fig:LinBilinGrads}.

In Fig. \ref{fig:distributions_abundances_to_iron} we investigate how abundance ratios relative to iron vary with Galactocentric radius for stars with reliable abundance measurements. Following Fig. \ref{fig:Gradients}, we present both a simple linear fit and a GP fit to the running median of the distribution. We find statistically significant positive gradients for [O/Fe], [Mg/Fe], and [Ca/Fe], with slopes of $0.013\,\mathrm{dex\,kpc^{-1}}$, $0.007\,\mathrm{dex\,kpc^{-1}}$, and $0.005\,\mathrm{dex\,kpc^{-1}}$, respectively. The increasing trends in [Mg/Fe] and [Ca/Fe] agree in sign but are weaker than those reported by \citet{Genovali15}, who obtained $0.013\,\mathrm{dex\,kpc^{-1}}$ for [Mg/Fe] and $0.018\,\mathrm{dex\,kpc^{-1}}$ for [Ca/Fe]. For [O/Fe], \citet{Nunnari25} derived a pronounced positive gradient of $0.043\,\mathrm{dex\,kpc^{-1}}$ using the O I 777 nm triplet with NLTE abundances, whereas \citet{DaSilva23} found an almost flat gradient of $0.003\,\mathrm{dex\,kpc^{-1}}$ from the 6300 \AA [O I] line, weakly affected by NLTE effects. In contrast, the other $\alpha$-elements Si and Ti exhibit nearly flat trends.
 
Among the odd-$Z$ elements, [Na/Fe] also shows an increasing trend, whereas Al remains nearly flat, consistent with other CCs studies \citep[e.g.][]{Genovali15, Nunnari25}. The iron-peak elements display a more complex behavior: [Ni/Fe] is essentially constant with Galactocentric radius, while [Co/Fe] rises with a slope of $0.017\,\mathrm{dex\,kpc^{-1}}$. Similar behaviors for these species were reported by \citet{Luck18}. We measure a negative gradient in [Mn/Fe], implying that Mn is underabundant relative to Fe in the outer disk. The neutron-capture abundance ratios [Y/Fe], [Ba/Fe], and [Eu/Fe] all show positive correlations with radius, rising toward larger Galactocentric distances. For the $r$-process dominated element Eu, this increasing trend in [Eu/Fe] had previously been reported in CCs by \citet[][$0.025\,\mathrm{dex\,kpc^{-1}}$]{DaSilva16} and in open clusters by \citet[][$0.016\,\mathrm{dex\,kpc^{-1}}$]{Magrini23}. In contrast, results for the $s$-process element Y in CCs are mixed. \citet{DaSilva16} finds a flat [Y/Fe] gradient, whereas \citet{Luck18} reports a positive slope of $0.021\,\mathrm{dex\,kpc^{-1}}$ for [Y/Fe]. A complete list of slopes and intercepts for all elements, for both linear and bi-linear fits, is provided in Table \ref{tab:gradients}.

\section{Discussion}
\label{sec:discussion}

Regarding the [X/Fe] ratios across the Galactic disk, the positive trends for [O/Fe], [Mg/Fe] with $\rm{R}_{\rm{GC}}$ are consistent with studies of open clusters \citep{Donor2018, Casamiquela2019, CarbajoHijarrubia2024} and CCs \citep{Genovali15, Luck18}. However, other CC-based studies find a flat trend for Mg \citep[e.g.][]{Lemasle13, Nunnari25} and O \citep{DaSilva23}. A positive trend in [$\alpha$/Fe] has also been observed in simulations with radial variations in gas infall across the disk. In such scenario, \citet{Minchev2013, Minchev2014} found a [Mg/Fe] gradient of $0.027\,\mathrm{dex\,kpc^{-1}}$ for stars younger than 2 Gyr, which is higher than our reported gradient of $0.007\,\mathrm{dex\,kpc^{-1}}$.

Among the iron-peak group, flat abundance patterns such as [Ni/Fe] are expected, as these elements are produced through nucleosynthetic pathways similar to those of Fe. In contrast, our [$\rm{Cr}_{\rm{LTE}}$/$\rm{Fe}_{\rm{NLTE}}$] measurements show an offset of about $\sim -0.1$ dex relative to the GCE models, and a declining trend with $\rm{R}_{\rm{GC}}$. Similarly, [Mn/Fe] also shows a negative correlation with increasing $\rm{R}_{\rm{GC}}$, likely reflecting the metallicity dependence of thermonuclear supernova yields \citep{Badenes2008, KobayashiNomoto2009}, which becomes more evident in the outer disk where the mean metallicity is lower. [Co/Fe] shows a positive trend that is not observed in other CC studies, with the exception of \citet{Luck18}. We attribute this difference to the underestimation of [Co/Fe] at lower metallicities in LTE analyses \citep[see][]{storm2025}.

For the Odd-Z elements, the reported enhancements are well documented in the literature \citep{Daflon2003, Genovali15, Trentin24b, Nunnari25}. Specifically, [Na/Fe] and [Al/Fe] are consistently above GCE tracks and solar-scaled values. As mentioned, this enhancement likely arises, at least for Na, from processed material being brought to the surface by dredge-up. In particular, material processed through the NeNa cycle can enhance Na at the surface \citep{Cavallo1996}, producing [Na/Fe] enhancements of the order of $\sim 0.2$ dex. At the same time, CN reactions within the CNO cycle deplete C, with [C/Fe] decreasing by $\sim -0.3$ dex in Cepheids \citep[see][]{Takeda2013}. The magnitude of these enhancements are also affected by stellar rotation, especially for C \citep{Ekstrom2012, Anderson2014}. Yet, this effect cannot account for the [Na/Fe] increase at super-solar metallicity. 

\begin{figure}
	\includegraphics[width=\columnwidth]{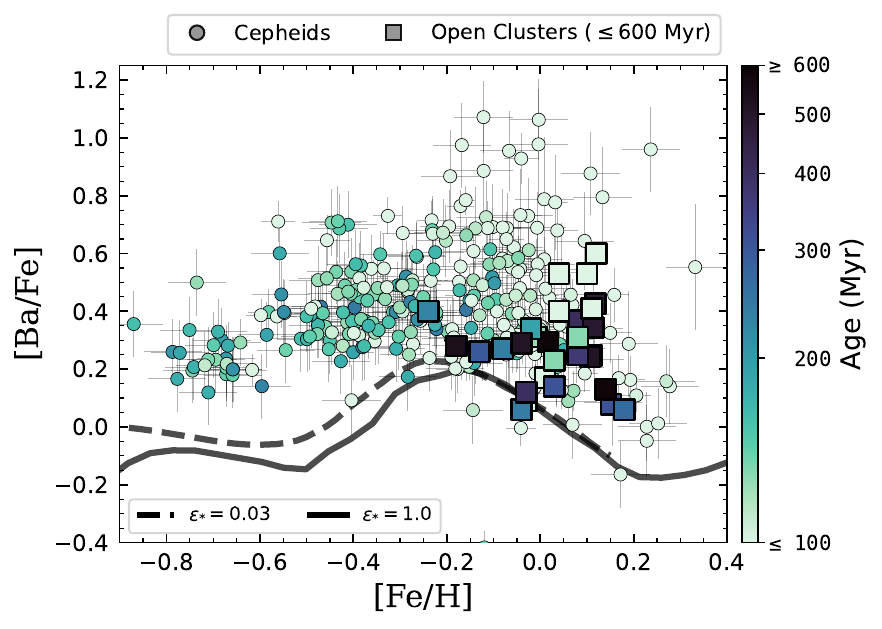}
    \caption{Barium abundance for Classical Cepheids and open clusters as a function of metallicity. Circles show CCs with ages estimated using the relations from \citet{Deng2025}. Squares show open clusters from \citet{Baratella2020, Baratella2021}, \citet{Magrini23} and \citet{CarbajoHijarrubia2024}. Black lines show the OMEGA+ GCE model from \citet{Lian2023} with $\epsilon_*$: 0.03 and $\epsilon_*$: 1.0 as in Fig. \ref{fig:distributions_abundances}.}
    \label{fig:Barium}
\end{figure}

As mentioned, among the s-process-dominated elements, [Ba/Fe] is the only abundance ratio that shows a peculiar bump around the solar metallicity. At metallicities $-0.7 \lesssim $ [Fe/H] $\lesssim -0.2$, the [Ba/Fe] ratio increases with [Fe/H], but a downturn occurs at [Fe/H]$\approx -0.2$ and then the ratio decreases to slightly over solar [Ba/Fe] around the solar [Fe/H] value. To further investigate this behavior of Ba, we compare our measurements with open clusters younger than 600 Myr, therefore spanning a similar age bracket to those of CCs. In Fig. \ref{fig:Barium}, we compare the distribution of [Ba/Fe] as a function of metallicity with open clusters from \citet{Baratella2020, Baratella2021}, \citet{CarbajoHijarrubia2024} and \citet{Magrini23}. For comparison, we also obtain ages for CCs using the Period-Age relation from \citet{Deng2025}. Overall, our measurements are compatible with those measured in open clusters of similar ages to those of CCs. 

The origin of this Ba enhancement in young stars is still unclear, and has been the subject of several studies. In open clusters, high barium abundance ratios, with [Ba/Fe] $\geq 0.5$ dex, and an increasing trend toward younger ages of $\sim 0.06 \, \rm{dex} \, Gyr^{-1}$ have been reported for stars in different evolutionary stages \citep{DOrazi2009, Yong2012, Jacobson2013, Baratella2021, CarbajoHijarrubia2024}. Similar behavior has also been observed in CCs by \citet{Andrievsky2013}, who found a nearly flat, super-solar [Ba/H] with galactocentric distance out to $15\,\mathrm{kpc}$. This qualitatively matches our results. The interpretation of this trend is further complicated by the fact that other heavy s-process elements, such as La, Ce, Nd, and Sm, do not show enhancements comparable to those of Ba \citep{DOrazi2017}.

The Ba enhancement has been partly attributed to limitations of stellar-atmosphere modeling, in particular the underestimation of \vmicnospace{} in LTE analyses \citep{Reddy2015, Reddy2017}. This LTE overestimation is further amplified as NLTE corrections for Ba are negative in stars with extended atmospheres \citep{Eitner2019}. However, as demonstrated here, our fully consistent NLTE analysis does not remove the anomalous enhancement. Moreover, elements with lines that also form at shallow atmospheric depths do not show comparable overabundances \citep{DOrazi2022}. We also note that the statistical uncertainties for Ba are $\sim 0.13$ dex and the phase to phase dispersion is 0.17 dex, significantly lower than the $\sim 1$ dex amplitude of the bump; therefore, it is unlikely to be an artifact.

From a purely nucleosynthetic perspective, the unusually high Ba abundances cannot be explained by the s-process alone \citep{Kapeller2011}. Additional contributions, such as those from the i-process \citep{Choplin2021}, are often invoked to reproduce the anomalous trends observed in open clusters \citep{Mishenina2015}. A more detailed investigation of this barium bump with GCE models would therefore be valuable, particularly models that include a more realistic star formation history and spatially varying star formation associated with spiral arms \citep[see][]{Spitoni2023}.

Among the other neutron-capture elements, the predominantly $r$-process element Eu shows a slight decrease in [Eu/Fe] with increasing [Fe/H] in our Cepheid sample, consistent with previous studies \citep{Battistini2016, storm2025} and the GCE model of \citet{Lian2023}. The derived ratios also overlap those of young open clusters \citep{VanderSwaelmen2023}. As a function of galactocentric radius, Y, Eu and Ba gradients relative to Fe show positive trends, broadly consistent with open clusters \citep{Magrini23, CarbajoHijarrubia2024} and CCs studies \citep{Yong2006, DaSilva16, Luck18}. For Eu, radial patterns may reflect the interplay between variations in star formation history and the enrichment timescales of $r$-process sources relative to those of Fe \citep{VanderSwaelmen2023, Molero2023}. In the case of [Y/Fe], the outer disk enhancement may be related to the metallicity dependence of yields from asymptotic giant branch stars \citep{Casali2020, Magrini2021}.

Our NLTE radial abundance gradients in the MW disk behave as expected, with elemental abundances steadily declining with increasing $R_{\rm GC}$. Beyond $R_{\rm GC} \sim 15\,\mathrm{kpc}$, the [X/H] gradients generally become shallower, in line with earlier work on Cepheids \citep{Luck18, Kovtyukh2022, DaSilva23, Nunnari25} and open clusters \citep{Magrini23}. The overall metallicity [M/H] follows the same pattern. The chemical enrichment picture from Milky Way H II regions, expected to be compatible with abundances of young stars, is less clear, with some studies finding evidence for outer disk flattening to be inconclusive \citep{fernandez2017}, whereas others argue against it \citep{Wang2018}. However, H II region abundance studies in other disk galaxies (e.g., M83 \citep{Bresolin2016}, NGC 3621 \citep{Kudritzki2014}, and NGC 1512 \citep{Bresolin2012}) do report flattening at large galactocentric radii, consistent with the overall trends we infer from CCs.

This absence of a tight correlation between gas-phase and stellar metallicity has been noted previously. \citet{MendezDelgado2024} showed that the relation between oxygen-based metallicities derived from star-forming regions and the actual Fe content of stars exhibits substantial intrinsic scatter, and is subject to systematic uncertainties such as depletion onto dust. This is further complicated by the substantial NLTE effects (see Fig. \ref{fig:NLTE_TSFitPy}) and their impact on radial oxygen gradients based on the O I 777 nm lines. In addition, \citet{Spitoni2023} found that elements like oxygen, produced on relatively short timescales, can show stronger azimuthal variations associated with spiral structure.

These considerations, which shape the detailed structure of the oxygen gradient, are particularly important for the distance ladder, since Cepheid metallicities are often extrapolated from gas-phase oxygen gradients in star-forming regions under the assumption of solar scaling \citep{Riess2022}. Overall, the consensus is that stellar metallicities broadly agree with those inferred from H II regions, though offsets of $\gtrsim 0.1$ dex can occur depending on the calibration adopted \citep[e.g.][]{Kudritzki24, Bresolin25}.

\section{Conclusions}
\label{sec:conclusions}

In this study, we present a new NLTE method to derive stellar parameters, iron metallicity, and abundance ratios for 16 other elements in pulsating stars, including CCs, Type II Cepheids, Anomalous Cepheids and RR Lyrae stars. The method is based on artificial neural networks and grids of synthetic stellar spectra with NLTE taken into account for 15 elements: Mg, Ca, Si, Ti, O, Mn, Fe, Ba, Eu, Y, Co, Sr, Na, Al, Ni and LTE for C and Cr. The method will be used to derive abundances based on high-resolution spectra of Cepheids to be acquired within the 4MOST 4MIDABLE-HR survey \citep{Bensby2019, S4paper} and here we outline the basic physical properties of the models, the methodology, and validate it on archival high-resolution spectra of Cepheids and other pulsating variables, to be observed by the 4MIDABLE-HR survey.

The code is validated using high-resolution UVES, HARPS, and FEROS spectra; the final catalog contains 351 Classical Cepheids, 33 Type II Cepheids, 8 Anomalous Cepheids, and 23 RR Lyrae stars. The cataloged CCs span a broad range in pulsation period, from 0.4 to 64.6 days, and a range of $-0.87$ to $+0.35$ dex in metallicity. We find that the basic atmospheric parameters of these stars can be obtained with typical uncertainties of 138 K for \teffnospace{}, 0.24 dex for $\log g$, and 0.06 dex for [Fe/H]. The average abundance uncertainties are 0.07 dex for the $\alpha$-elements O, Mg, Ca, Si and Ti; 0.07 dex for the odd-Z elements Na and Al; 0.17 dex for C; 0.07 dex for the iron-peak elements Cr, Mn, Co, and Ni; and 0.13 dex for the neutron-capture elements Y, Eu and Ba, with Sr having a higher uncertainty of 0.19 dex.

We carry out a multi-step validation of the NLTE spectroscopic results in the framework of stellar structure modeling. For the 23 calibrators with at least five accepted spectra, the mean per-star standard deviation across the sampled pulsation phases is 0.07 dex in [Fe/H]. The average scatter in the other reported [X/Fe] abundance ratios is below 0.1 dex, with the exception of Co, Ba, and Sr, which show mean dispersions of 0.11, 0.17, and 0.14 dex, respectively. Comparing the spectroscopic \teffnospace{} and luminosities with expectations from stellar evolution models for classical pulsators \citep{Bono99, Anderson16} reveals excellent agreement at solar metallicity, and the spectroscopic models provide consistent solutions for phase dependent variations in atmospheric parameters over the pulsation cycle \citep{Vasilyev18}. Among the 23 multiphase stars, the radius changes driven by pulsation range from 2\% to 18\% of the mean stellar radius. The derived changes in \teffnospace{} and \loggnospace{} are consistent with the location of the instability strip and with theoretical predictions for pulsating stars \citep{Anderson16}.

We use our NLTE abundances together with distances to trace chemical evolution as a function of [Fe/H] and to derive NLTE radial abundance gradients in the Milky Way disk. Our [Fe/H] and [M/H] fits indicate a change in radial slope near $R_{\rm GC} \sim 15\,\mathrm{kpc}$, consistent with previous work, although the outer slopes remain poorly constrained. The [X/Fe]–[Fe/H] relations for the elements in our sample generally support the GCE predictions from \citet{Lian2023} based on FGKM-type stars. Discrepancies with the GCE models remain for Al and Ti, as previously reported in the literature, while the expected departures for C and Na reflect stellar evolutionary effects. The radial [X/Fe] trends, including elevated [O/Fe], [Mg/Fe], and [Ca/Fe] in the outer disk, are consistent with GCE models involving radial variations in gas infall \citep{Minchev2013, Minchev2014}. Among the iron-peak elements, Ni remains nearly constant, whereas [Co/Fe] displays a rising trend, likely due to NLTE effects becoming more pronounced at lower metallicities. Both [Cr/Fe] and [Mn/Fe] decrease with galactocentric radius. In contrast, neutron-capture elements appear to be enhanced relative to iron toward the outer regions of the Galaxy.

We also report a peculiar bump structure in our NLTE Ba abundances around solar metallicity. The origin of this [Ba/Fe] enhancement is not known, but it resembles trends reported for FGKM-type stars \citep{Storm2026}. This trend is also consistent with the [Ba/Fe] behavior observed in young open clusters of similar age \citep{DOrazi2017, Baratella2021, CarbajoHijarrubia2024}.

We present NLTE radial abundance gradients for several key elements, including for the first time the NLTE trends with R$_{\rm{GC}}$ of Co, Ni, Y, and Eu for Cepheids. We find a decrease in abundances with galactocentric distance consistent with models and observations from MW-like galaxies.

The method presented here will be used to analyze Classical Cepheids and other types of pulsators in the 4MIDABLE-HR survey. Accurate chemical abundances of pulsating stars are important for a wide range of Galactic and extragalactic studies, and the 4MIDABLE-HR dataset will provide an unprecedented sample size and data quality, enabling new constraints on chemical gradients in the Milky Way. An improved understanding of the relationships between pulsation properties and atmospheric parameters of Cepheids, especially with new low-metallicity data from 4MIDABLE-HR, will provide critical insight into the dependence of the fundamental period-luminosity relation on metallicity and individual abundances. This will help improve the precision of the extragalactic distance calibration, which is of great relevance to modern cosmology.

\begin{acknowledgements}
We sincerely thank the referee for the positive feedback and comments that improved the readability of the paper. RA, NS, MB and GG acknowledge funding from the European Research Council (ERC) under the European Union’s Horizon 2020 research and innovation programme (Grant agreement No. 949173). MB is supported through the Lise Meitner grant from the Max Planck Society. We acknowledge support by the Collaborative Research Centre SFB 881 (projects A5, A10), Heidelberg University, of the Deutsche Forschungsgemeinschaft (DFG, German Research Foundation). GG acknowledges support by Deutsche Forschungs-gemeinschaft (DFG, German Research Foundation) – project-IDs: eBer-22-59652 (GU 2240/1-1  "Galactic Archaeology with Convolutional Neural-Networks: Realising the  potential of Gaia and 4MOST"). RIA is funded by zukunft.niedersachsen, the joint science funding program of the Lower Saxony Ministry of Science and Culture and the Volkswagen Foundation. VD acknowledges partial support from INAF Minigrant 2024 MUGS. T.B. acknowledges financial support by grant No.~2024-04990 from the Swedish Research Council. GT acknowledges financial support of the Slovenian Research Agency (research core funding No. P10188) and the European Space Agency (Prodex Experiment Arrangement No. 4000143450).

\end{acknowledgements}

\bibliographystyle{aa}
\bibliography{mybib}

\begin{appendix}
\nolinenumbers

\section{Modeling individual lines}
\label{app:lines}

Due to the pulsating nature of variables, fundamental parameters are expected to change over the period \citep{Bono99, Luck18}. This not only impacts the \teffnospace{} and \loggnospace{}, but also the turbulent velocities $V_{\rm{mic}}$ and $V_{\rm{mac}}$ \citep{Aerts2014, Proxauf18}. Due to these changes, line profiles change accordingly both in depth and width. This can be observed for a multi-phase star shown in the top panel of Fig. \ref{fig:CalibMuliphase}. The combination of such changes can significantly impact the line profiles (See the bottom left panel of Fig. \ref{fig:CalibMuliphase}), making lines potentially unsuitable for reliable abundances depending on the stellar parameters and the phase of observation. To filter such cases, we used the \texttt{iSpec} routines \citep{Blanco-Cuaresma14} to model the profiles and depth of individual lines considered in this study. We discard lines that are not well represented by a simple Gaussian profile and have a minimum depth of under 3\% of the continuum. This allows us to discard lines with significant blending or heavily influenced by continuum placement. The full list of lines used in this study can be seen in Table \ref{tab:lines}.

\begin{table}[!htbp]
\caption{Atomic data of the lines used in the fitting of different elements.\label{tab:lines}}
\centering
\scriptsize
\setlength{\tabcolsep}{3pt}
\resizebox{\columnwidth}{!}{\begin{tabular}{lccccc}
\hline\hline
Element & $\lambda$ [\AA] & $E_{\rm low}$ [eV] & $\log(gf)$ & $g_{\rm up}$ & Isotope \\
\hline \vspace{1pt}
Al I  & 6696.023 & 3.143 & -1.569 & 4 &      \\
Al I  & 6698.673 & 3.143 & -1.870 & 2 &      \\
Ba II & 6141.709 & 0.704 & -0.395 & 4 & 137  \\
Ba II & 6141.713 & 0.704 & -0.032 & 4 & 134  \\
Ba II & 6141.714 & 0.704 & -0.032 & 4 & 136  \\
\ldots & \ldots & \ldots & \ldots & \ldots &  \\
Y II & 5662.924 & 1.944 &  0.340 & 7 \\ \hline
\end{tabular}}
\end{table}

\section{Atmospheric parameter validation}
\label{app:biases}

\begin{figure}
	\includegraphics[width=\columnwidth]{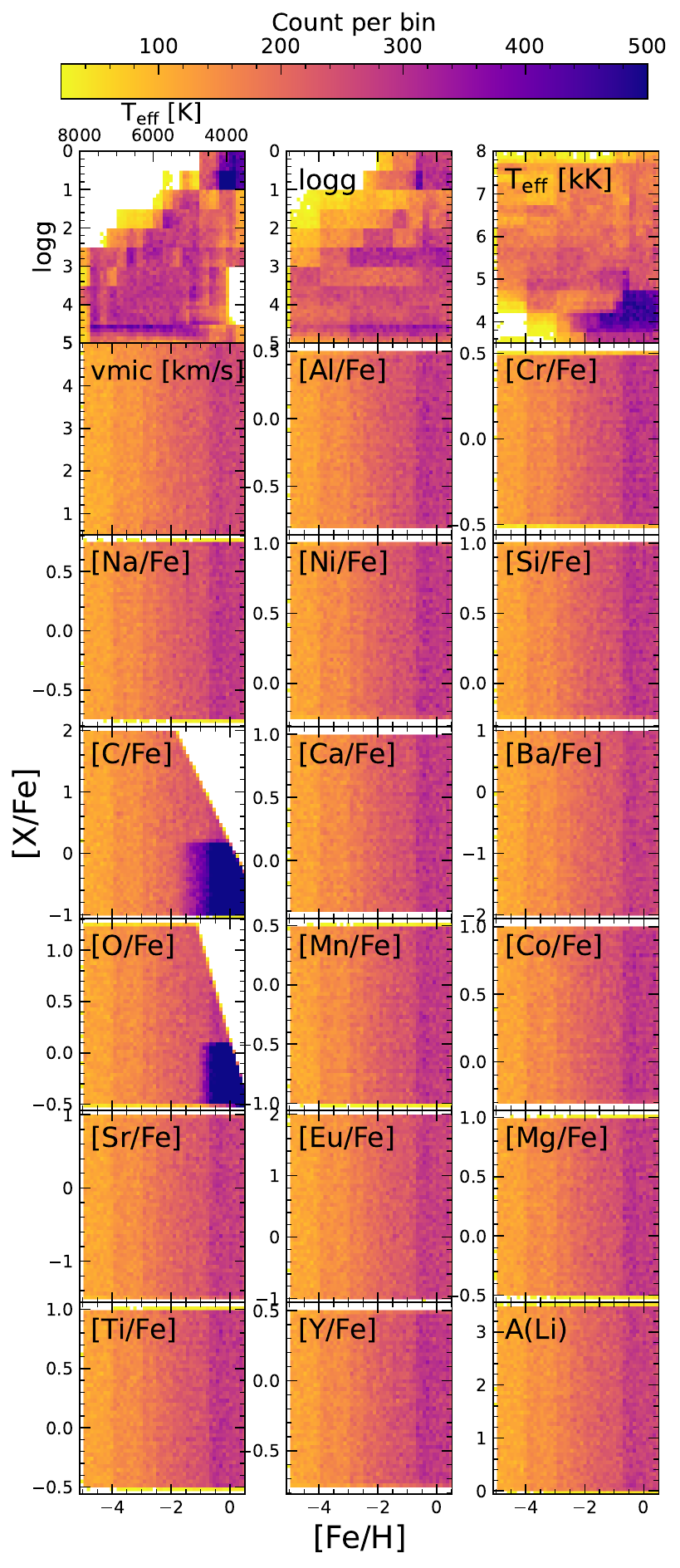}
    \caption{Distribution in a 2D histogram of synthetic spectra used to train the Payne ANN-emulator. Top left panel shows the Kiel diagram, while the rest are distributions as a function of [Fe/H]. All abundances are chosen uniformly random in metallicity space, except for A(O)$< 8.87$ and A(C)$< 8.7$. There are fewer low metallicity giant model atmospheres, resulting in slightly fewer spectra at low metallicities.}
    \label{fig:paramspace}
\end{figure}

\begin{figure}[tbp]
	\includegraphics[width=\columnwidth]{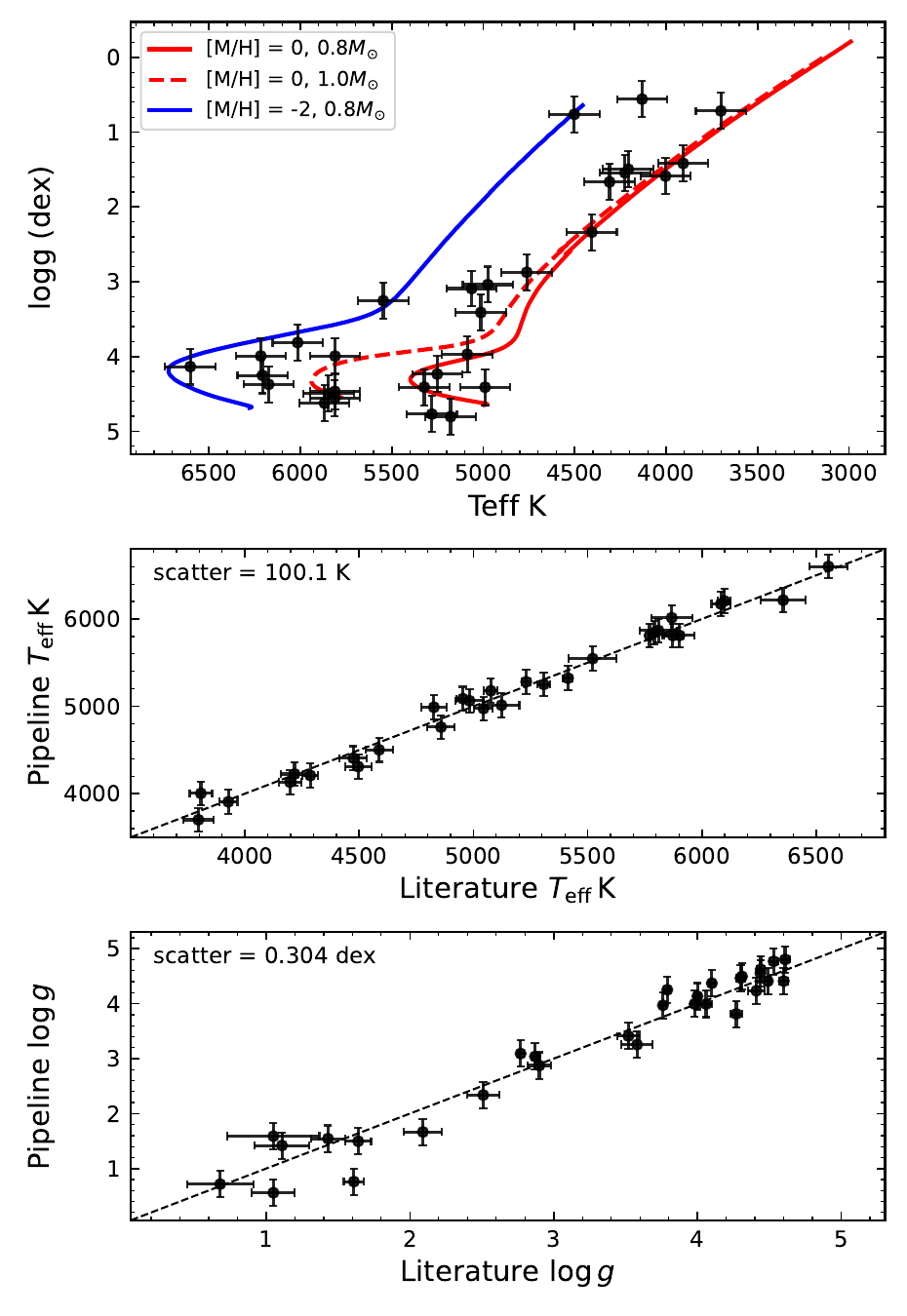}
    \caption{Comparison between the literature atmospheric parameters of \teffnospace{} and \loggnospace{} for benchmark stars in \citet{Heiter2015b}. Top panel shows the Kiel diagram for benchmark stars derived with our method. Colors show the PARSEC evolutionary tracks \citep{Bressan2012} for [M/H] = 0 (red) and [M/H] = -2 (blue).  Middle and bottom panels show the comparison with \citet{Heiter2015b} in \teffnospace{} and \loggnospace{} for their reported values (x-axis) and our pipeline (y-axis), respectively after a bias correction of -30 K in \teffnospace{} and -0.1 in \loggnospace{}. The standard deviations of the residuals for both cases are shown on the top left.}
    \label{fig:APP_biases}
\end{figure}

In this section, we show the distribution of synthetic spectra used to train the ANN-emulator. The complete parameter space is shown in Fig. \ref{fig:paramspace}. We fit the archival spectra compiled for this sample and derive \teffnospace{} and \loggnospace{} with our method. We then compare these results with the \teffnospace{} and \loggnospace{} reported in \citet{Heiter2015b}, as shown in Fig. \ref{fig:APP_biases}.

We then validate the derived abundances by fitting the solar spectra. In Table \ref{tab:all_solar}, we show the average abundances obtained from the seven solar spectra observed with HARPS, UVES, NARVAL and FEROS. We use these results to compute stellar abundances in square-bracket notation by correcting the abundance derived for a given spectrum for the difference between our solar values and those of \citet{Magg2022}. We note that the largest offsets correspond to C, Ba, Y, and Eu. The Y and Eu offsets are likely due to the accuracy of the $\log gf$ values used in the line list \citep[see][]{Storm2024, Storm2026}. The oxygen abundance scatter is mainly driven by the single NARVAL spectrum that yielded [O/Fe] $= -0.33$; however, there is no clear indication of problems with the observed spectrum, and we therefore included it in the calculation. 

\begin{table}[!htbp]
\centering
\caption{Averaged fitted solar abundances from 7 4MOST-like solar spectra.}
\label{tab:all_solar}
\begin{tabular}{cc}
\toprule
Element &  Average fitted abundance  \\ \hline
\midrule
C  & $8.34 \pm 0.10$ \\
O  & $8.68 \pm 0.15$ \\
Na & $6.22 \pm 0.02$ \\
Mg & $7.39 \pm 0.03$ \\
Al & $6.40 \pm 0.03$ \\
Si & $7.52 \pm 0.01$ \\
Ca & $6.33 \pm 0.02$ \\
Ti & $4.96 \pm 0.02$ \\
Cr & $5.61 \pm 0.04$ \\
Mn & $5.39 \pm 0.02$ \\
Co & $4.97 \pm 0.03$ \\
Ni & $6.28 \pm 0.03$ \\
Sr & $2.88 \pm 0.08$ \\
Y  & $2.08 \pm 0.04$ \\
Ba & $2.07 \pm 0.03$ \\
Eu & $0.80 \pm 0.11$ \\

\hline
\bottomrule
\end{tabular}
\end{table}

\FloatBarrier

\onecolumn
\raggedbottom

\begin{figure}[H]
\centering
	\includegraphics[width=\textwidth]{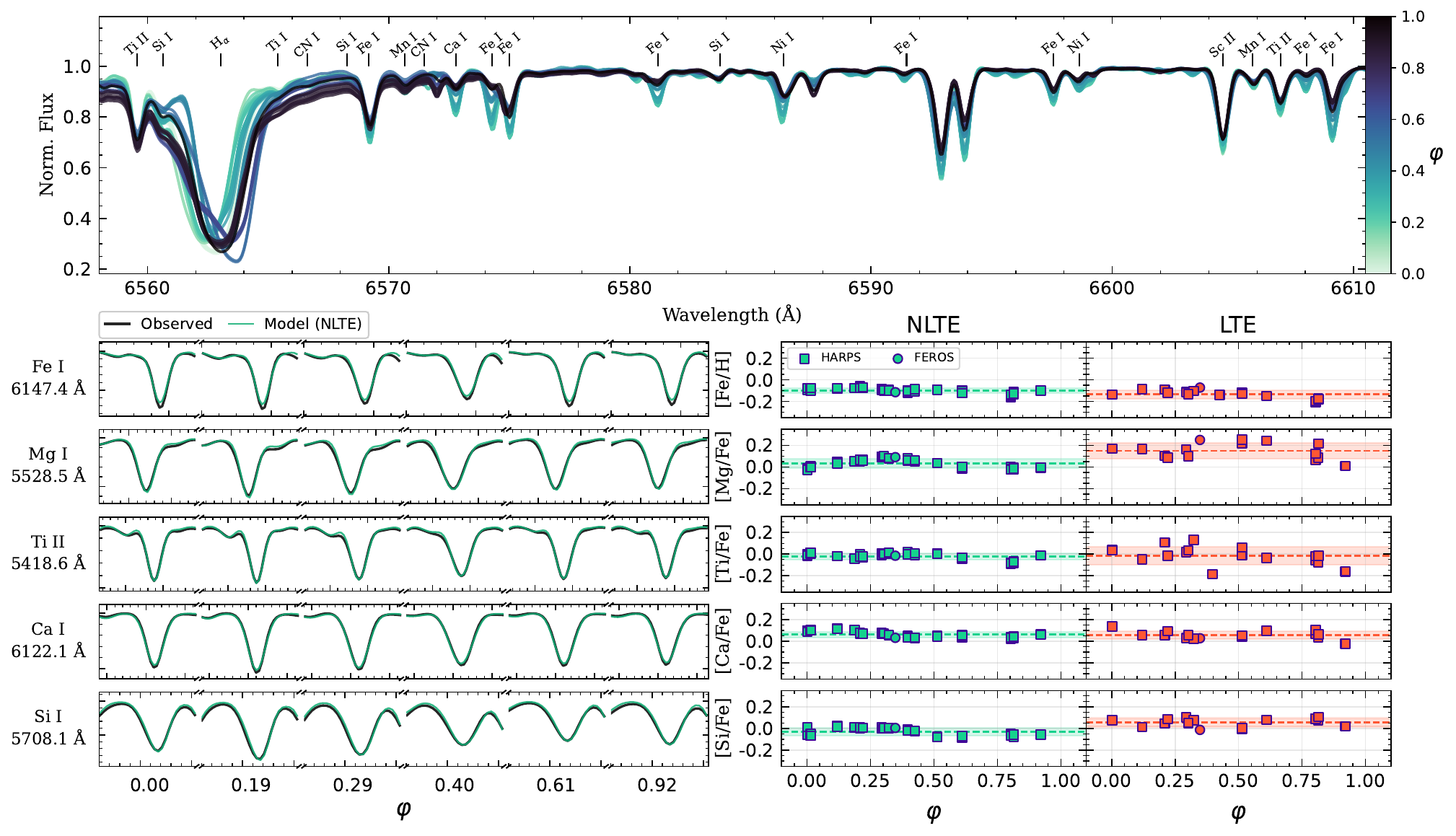}
    \caption{Same as Fig. \ref{fig:CalibMuliphase} for the star $\beta$-Dor.}
    \label{fig:CalibMuliphaseBetaDor}
\end{figure}

\begin{figure}[H]
\centering
	\includegraphics[width=\textwidth]{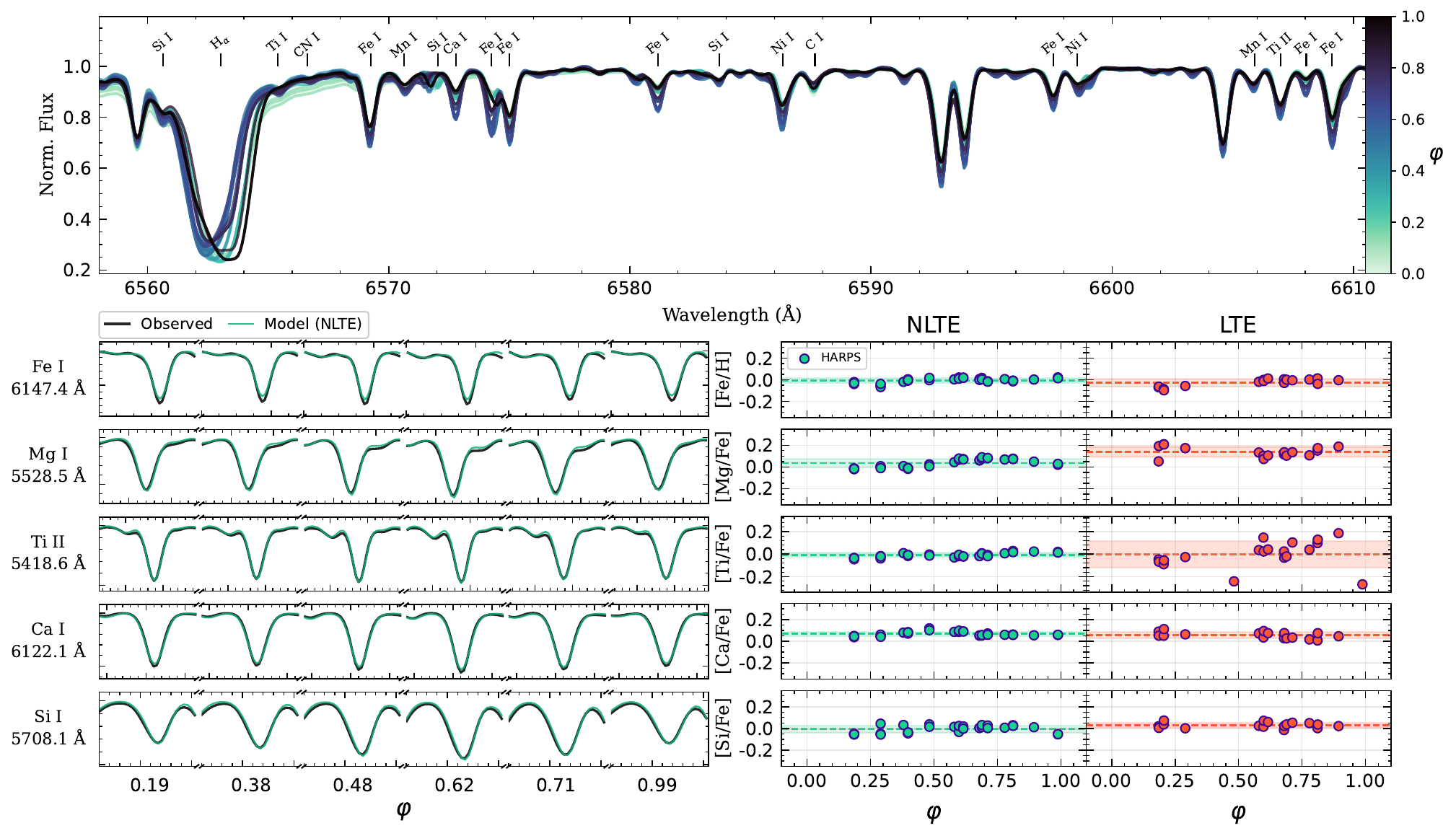}
    \caption{Same as Fig. \ref{fig:CalibMuliphase} for the star $\zeta$-Gem.}
    \label{fig:CalibMuliphaseZetaGem}
\end{figure}

\begin{figure}[H]
\centering
	\includegraphics[width=\textwidth]{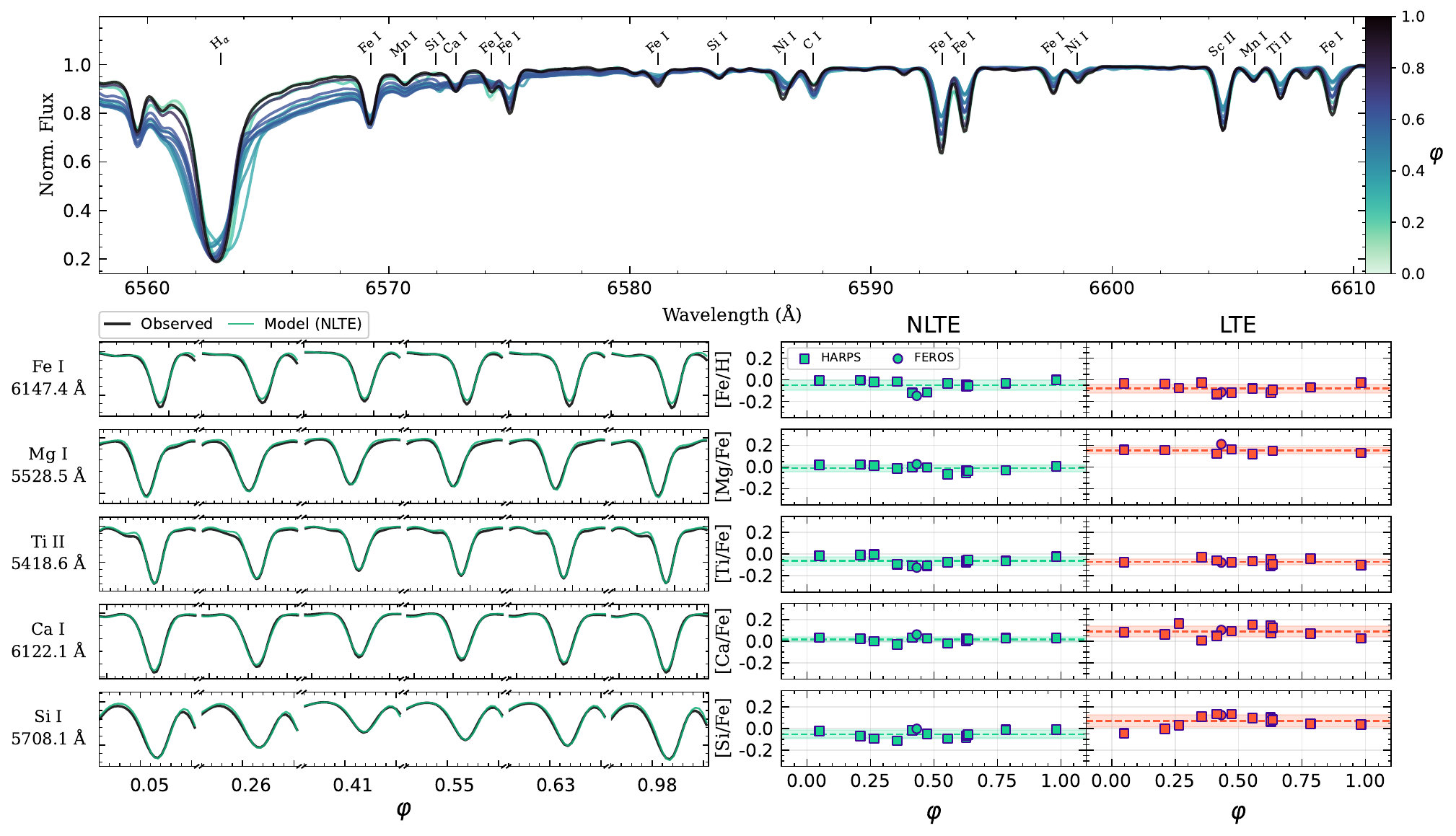}
    \caption{Same as Fig. \ref{fig:CalibMuliphase} for the star S-Cru.}
    \label{fig:CalibMuliphaseRTra}
\end{figure}

\begin{figure}[H]
\centering
	\includegraphics[width=\textwidth]{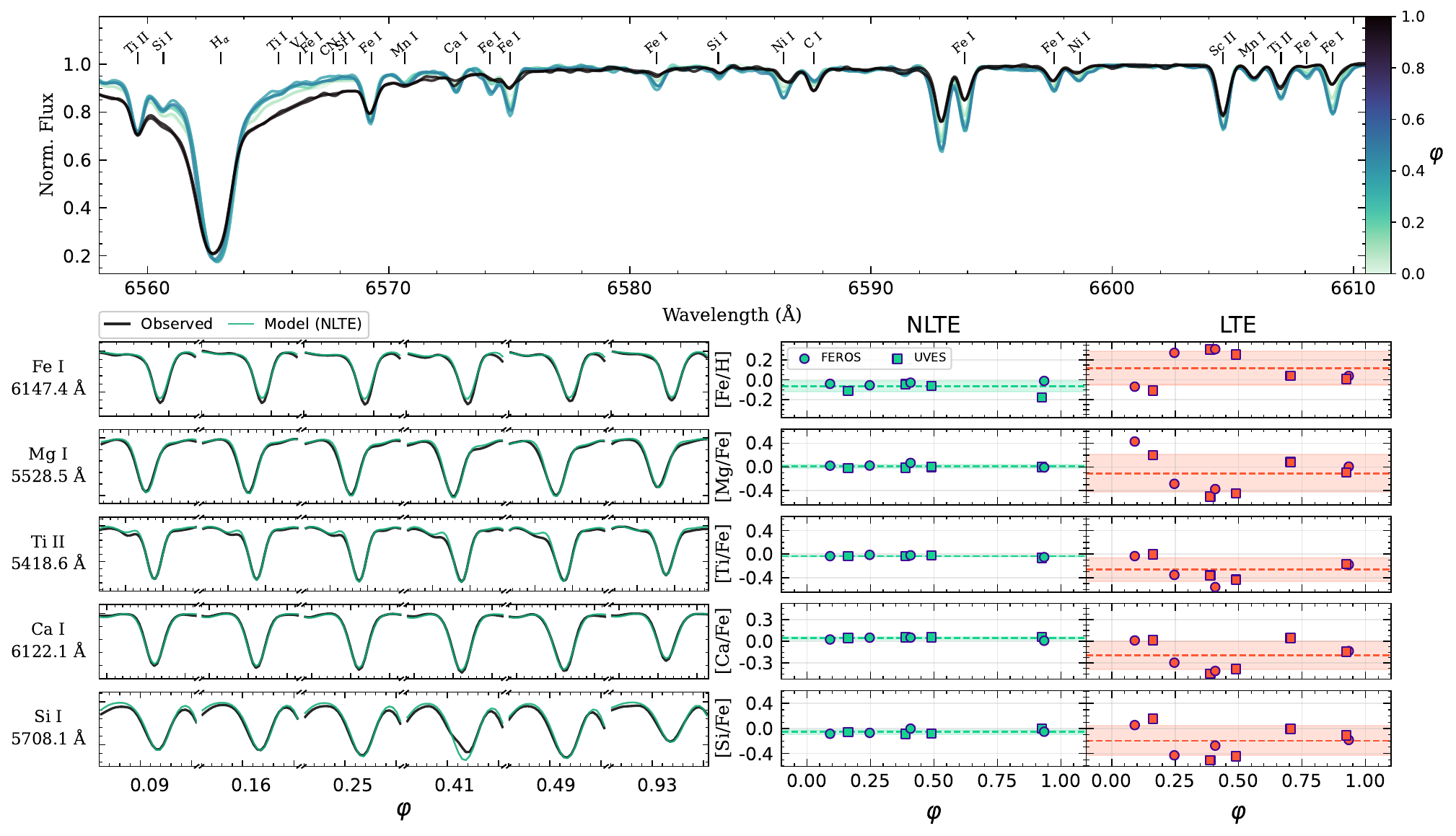}
    \caption{Same as Fig. \ref{fig:CalibMuliphase} for the star XX-Sgr.}
    \label{fig:CalibMuliphaseXXSgr}
\end{figure}

\begin{figure}[H]
\centering
	\includegraphics[width=\textwidth]{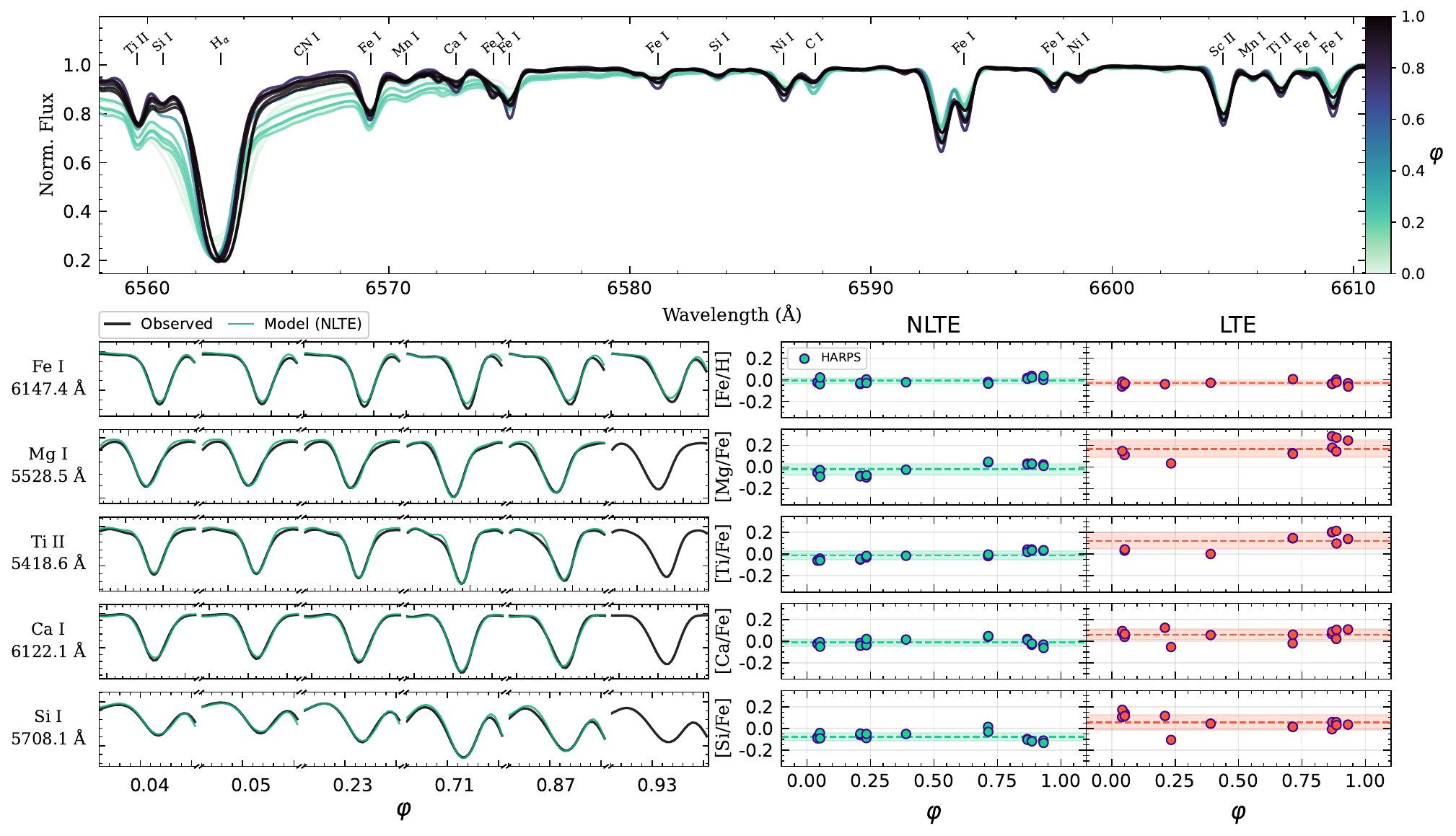}
    \caption{Same as Fig. \ref{fig:CalibMuliphase} for the star Y-Sgr.}
    \label{fig:CalibMuliphaseYSgr}
\end{figure}

\noindent\begin{minipage}[t]{0.48\textwidth}
\vspace{0pt}
\section{Multiphase abundances}

\label{app:MultiphaseAll}

In this section, we present the multi-phase abundances in NLTE and LTE and the changes in line profiles for the remaining example stars with ten or more individual spectra across pulsation in Fig \ref{fig:CalibMuliphaseBetaDor}, Fig. \ref{fig:CalibMuliphaseZetaGem}, Fig. \ref{fig:CalibMuliphaseRTra}, Fig. \ref{fig:CalibMuliphaseXXSgr} and Fig. \ref{fig:CalibMuliphaseYSgr}. 

We also show the average abundance $1\sigma$ dispersions obtained with our method for all target elements across the 23 stars with five or more spectra across pulsation. We calculate the standard deviation across phase for each element in all of the 23 stars with multiphase observations, we then average the $1\sigma$ dispersions for all elements and report the results in Table \ref{tab:abundance_dispersion}.

\end{minipage}\hfill\begin{minipage}[t]{0.48\textwidth}
\vspace{0pt}\setlength{\intextsep}{0pt}
\begin{table}[H]
\caption{Mean abundance dispersions as a function of phase for the multiphase sample of 23 stars. The values represent the average of standard deviations among the multiphase stars, along with the standard deviations associated with those averages.}

\label{tab:abundance_dispersion}
\centering
\begin{tabular}{lc}
\hline
Abundance & Average standard deviation (dex) \\
\hline
$[\mathrm{C/Fe}]$  & $0.10 \pm 0.09$ \\
$[\mathrm{O/Fe}]$  & $0.06 \pm 0.05$ \\
$[\mathrm{Na/Fe}]$ & $0.04 \pm 0.03$ \\
$[\mathrm{Mg/Fe}]$ & $0.07 \pm 0.06$ \\
$[\mathrm{Al/Fe}]$ & $0.05 \pm 0.05$ \\
$[\mathrm{Si/Fe}]$ & $0.05 \pm 0.04$ \\
$[\mathrm{Ca/Fe}]$ & $0.03 \pm 0.02$ \\
$[\mathrm{Ti/Fe}]$ & $0.06 \pm 0.07$ \\
$[\mathrm{Cr/Fe}]$ & $0.04 \pm 0.03$ \\
$[\mathrm{Mn/Fe}]$ & $0.08 \pm 0.07$ \\
$[\mathrm{Fe/H}]$  & $0.07 \pm 0.06$ \\
$[\mathrm{Ni/Fe}]$ & $0.04 \pm 0.05$ \\
$[\mathrm{Sr/Fe}]$ & $0.14 \pm 0.13$ \\
$[\mathrm{Y/Fe}]$  & $0.04 \pm 0.02$ \\
$[\mathrm{Ba/Fe}]$ & $0.17 \pm 0.15$ \\
$[\mathrm{Eu/Fe}]$ & $0.07 \pm 0.10$ \\
$[\mathrm{Co/Fe}]$ & $0.11 \pm 0.10$ \\
\hline
\end{tabular}
\tablefoot{This table is available in machine-readable form in the online article.}
\end{table}
\end{minipage}

\clearpage

\clearpage

\clearpage
\begingroup\small
\setlength{\columnsep}{2em}
\begin{multicols}{2}
\setlength{\abovedisplayskip}{3pt}\setlength{\belowdisplayskip}{3pt}\setlength{\abovedisplayshortskip}{3pt}\setlength{\belowdisplayshortskip}{3pt}
\section{Gradients}
\label{app:gradients}

For each abundance ratio, we fit three different models: a linear model, a broken-linear model and a Gaussian process (GP) model. The linear and broken linear fits are obtained using a maximum likelihood method that accounts both for uncertainties in abundances and distances together. For each CC in our sample, $\rm{CC}_{i}$, with measured galactocentric radius $R_{\mathrm{GC},i}^{\mathrm{obs}}$ and abundance denoted by $y_i=[\mathrm{X/H}]_i$ or $y_i= [\mathrm{X/Fe}]_i$, with their respective uncertainties $\sigma_{\rm{R}_{\rm{GC, i}}}$ and $\sigma_{y, i}$, we use the latent coordinate $\rm{R}_{i}$, so the likelihood can be written as

\begin{equation}
\begin{aligned}
\mathcal{L}\!\left(
\boldsymbol{\theta},s_{\mathrm{int}},\{R_i\}
\mid \{R_{\mathrm{GC},i}^{\mathrm{obs}},y_i\}
\right)
&=
\prod_{i=1}^{N}
\mathcal{N}\!\left(
R_{\mathrm{GC},i}^{\mathrm{obs}}
\mid R_i,\sigma_{R_{\mathrm{GC}},i}^{2}
\right)
\\
&\, \, \, \quad \qquad\times
\mathcal{N}\!\left(
y_i
\mid f(R_i;\boldsymbol{\theta}),
\sigma_{y,i}^{2}+s_{\mathrm{int}}^{2}
\right).
\end{aligned}
\label{eq:eiv_likelihood}
\end{equation}

Here, $s_{\rm int}$ represents the intrinsic scatter, and $f(X, \theta)$ is, in this case a linear or broken-linear relation parameterized by $\theta$. The term $\mathcal{N}(z\mid\mu,\sigma^2)$ denotes a normal probability density with mean $\mu$ and variance $\sigma^2$. Parameter uncertainties are estimated from the standard deviations of the bootstrap distributions, using $10^{4}$ resamples for the simple linear fits. For bi-linear fits, we instead report the 16th–84th bootstrap percentiles from the $10^{4}$ resamples. This choice is motivated by outliers arising from the smaller outer-disk sample, which produce heavy tails in the bootstrap distributions.

Some of the bilinear fits on single elements such as Na, Mn, Co and Ni locate the break radius inside the solar radius with very steep inner gradients. As we chose against including constraints or priors in the fit, some elements with comparatively fewer stars such as Mn, Co and Ni may produce spurious features in their fits. We include these results in Table \ref{tab:gradients}, however these results are not as robust as [Fe/H] and [M/H].

In the case of the GP regression, we adopt a flexible kernel composed of a radial basis function, a constant kernel, and a white-noise kernel, fitted to the running median, written as, 
\begin{equation}
K_{ij} = \sigma_f^2 \exp\left[ -\frac{(x_i-x_j)^2}{2\ell^2} \right]
+
\left( \sigma_{\mathrm{w}}^2+\sigma_{y_i}^{\,2} \right)\delta_{ij}.
\end{equation}

This approach is intended to capture variations in Cepheid abundances across the Galactic disk without imposing strong assumptions on the underlying shape of the chemical trends. The results for the linear fitting for all target elemental abundances, relative to H and Fe, are shown in Table \ref{tab:gradients}.

\end{multicols}
\endgroup
\begin{table}[H]
\centering
\scriptsize
\caption{Radial abundance gradients.}
\label{tab:gradients}
\renewcommand{\arraystretch}{0.90}
\setlength{\tabcolsep}{3.5pt}
\providecommand{\gradientpm}[2]{\makebox[3.3em][r]{$#1$}\ensuremath{{}\pm{}}\makebox[3.3em][l]{$#2$}}
\providecommand{\gradientci}[3]{\makebox[3.3em][r]{$#1$}\ensuremath{{}_{-#2}^{+#3}}}
\begin{tabular*}{\linewidth}{@{\extracolsep{\fill}}lcccc@{\hspace{1.8em}}lcccc@{}}
\hline\noalign{\vskip2pt}
\multicolumn{5}{c}{Linear fits: $[\mathrm{X}/\mathrm{H}]$, $[\mathrm{M}/\mathrm{H}]$} & \multicolumn{5}{c}{Linear fits: $[\mathrm{X}/\mathrm{Fe}]$} \\[2pt]
\cline{1-5}\cline{6-10}\noalign{\vskip2pt}
Ratio & Slope $m$ & Zero-point $b$ & $s_{\mathrm{int}}$ & $N_{\mathrm{stars}}$ & Ratio & Slope $m$ & Zero-point $b$ & $s_{\mathrm{int}}$ & $N_{\mathrm{stars}}$ \\[1pt]
 & (dex/kpc) & (dex) & (dex) & & & (dex/kpc) & (dex) & (dex) & \\[2pt]
\hline
$[\mathrm{C}/\mathrm{H}]$ & $-0.046 \pm 0.003$ & $0.187 \pm 0.043$ & $0.000$ & 303 & $[\mathrm{C}/\mathrm{Fe}]$ & $0.003 \pm 0.004$ & $-0.135 \pm 0.053$ & $0.000$ & 303 \\
$[\mathrm{O}/\mathrm{H}]$ & $-0.039 \pm 0.003$ & $0.210 \pm 0.029$ & $0.067$ & 329 & $[\mathrm{O}/\mathrm{Fe}]$ & $0.013 \pm 0.002$ & $-0.169 \pm 0.020$ & $0.079$ & 329 \\
$[\mathrm{Na}/\mathrm{H}]$ & $-0.043 \pm 0.002$ & $0.538 \pm 0.025$ & $0.110$ & 322 & $[\mathrm{Na}/\mathrm{Fe}]$ & $0.006 \pm 0.002$ & $0.188 \pm 0.018$ & $0.066$ & 322 \\
$[\mathrm{Mg}/\mathrm{H}]$ & $-0.043 \pm 0.002$ & $0.330 \pm 0.025$ & $0.094$ & 330 & $[\mathrm{Mg}/\mathrm{Fe}]$ & $0.007 \pm 0.001$ & $-0.028 \pm 0.015$ & $0.057$ & 330 \\
$[\mathrm{Al}/\mathrm{H}]$ & $-0.052 \pm 0.003$ & $0.496 \pm 0.032$ & $0.087$ & 217 & $[\mathrm{Al}/\mathrm{Fe}]$ & $0.003 \pm 0.002$ & $0.098 \pm 0.019$ & $0.000$ & 217 \\
$[\mathrm{Si}/\mathrm{H}]$ & $-0.056 \pm 0.003$ & $0.393 \pm 0.028$ & $0.080$ & 259 & $[\mathrm{Si}/\mathrm{Fe}]$ & $-0.003 \pm 0.001$ & $0.003 \pm 0.014$ & $0.029$ & 259 \\
$[\mathrm{Ca}/\mathrm{H}]$ & $-0.047 \pm 0.002$ & $0.373 \pm 0.023$ & $0.082$ & 334 & $[\mathrm{Ca}/\mathrm{Fe}]$ & $0.005 \pm 0.001$ & $-0.004 \pm 0.010$ & $0.000$ & 334 \\
$[\mathrm{Ti}/\mathrm{H}]$ & $-0.048 \pm 0.003$ & $0.311 \pm 0.027$ & $0.098$ & 334 & $[\mathrm{Ti}/\mathrm{Fe}]$ & $0.003 \pm 0.001$ & $-0.047 \pm 0.014$ & $0.000$ & 334 \\
$[\mathrm{Cr}/\mathrm{H}]$ & $-0.057 \pm 0.002$ & $0.369 \pm 0.026$ & $0.090$ & 335 & $[\mathrm{Cr}/\mathrm{Fe}]$ & $-0.008 \pm 0.001$ & $0.029 \pm 0.014$ & $0.000$ & 335 \\
$[\mathrm{Mn}/\mathrm{H}]$ & $-0.053 \pm 0.003$ & $0.489 \pm 0.029$ & $0.118$ & 263 & $[\mathrm{Mn}/\mathrm{Fe}]$ & $-0.010 \pm 0.001$ & $0.175 \pm 0.016$ & $0.052$ & 263 \\
$[\mathrm{Fe}/\mathrm{H}]$ & $-0.051 \pm 0.002$ & $0.371 \pm 0.024$ & $0.131$ & 335 & \multicolumn{5}{c}{---} \\
$[\mathrm{Co}/\mathrm{H}]$ & $-0.020 \pm 0.004$ & $0.300 \pm 0.037$ & $0.137$ & 181 & $[\mathrm{Co}/\mathrm{Fe}]$ & $0.017 \pm 0.003$ & $0.018 \pm 0.025$ & $0.083$ & 181 \\
$[\mathrm{Ni}/\mathrm{H}]$ & $-0.043 \pm 0.003$ & $0.203 \pm 0.037$ & $0.000$ & 262 & $[\mathrm{Ni}/\mathrm{Fe}]$ & $-0.001 \pm 0.001$ & $-0.083 \pm 0.014$ & $0.000$ & 262 \\
$[\mathrm{Y}/\mathrm{H}]$ & $-0.042 \pm 0.003$ & $0.240 \pm 0.037$ & $0.000$ & 335 & $[\mathrm{Y}/\mathrm{Fe}]$ & $0.008 \pm 0.001$ & $-0.093 \pm 0.012$ & $0.000$ & 335 \\
$[\mathrm{Ba}/\mathrm{H}]$ & $-0.047 \pm 0.004$ & $0.718 \pm 0.048$ & $0.232$ & 296 & $[\mathrm{Ba}/\mathrm{Fe}]$ & $0.004 \pm 0.003$ & $0.355 \pm 0.042$ & $0.166$ & 296 \\
$[\mathrm{Eu}/\mathrm{H}]$ & $-0.035 \pm 0.002$ & $0.074 \pm 0.025$ & $0.102$ & 327 & $[\mathrm{Eu}/\mathrm{Fe}]$ & $0.017 \pm 0.001$ & $-0.306 \pm 0.013$ & $0.000$ & 327 \\
$[\mathrm{M}/\mathrm{H}]$ & $-0.044 \pm 0.002$ & $0.277 \pm 0.022$ & $0.098$ & 335 & \multicolumn{5}{c}{---} \\
\hline
\end{tabular*}
\par\vspace{0.9em}
\renewcommand{\arraystretch}{1.00}
\begin{tabular*}{\linewidth}{@{\extracolsep{\fill}}lccccccc@{}}
\hline\noalign{\vskip2pt}
\multicolumn{8}{c}{Broken-linear fits: $[\mathrm{X}/\mathrm{H}]$, $[\mathrm{M}/\mathrm{H}]$} \\[2pt]
\hline\noalign{\vskip2pt}
Ratio & Inner slope $m_1$ & Outer slope $m_2$ & Break radius $R_{\mathrm{b}}$ & Abundance at break $y_{\mathrm{b}}$ & $s_{\mathrm{int}}$ & $N_{\mathrm{stars}}$ & $N_{\mathrm{break}}$ \\[1pt]
 & (dex/kpc) & (dex/kpc) & (kpc) & (dex) & (dex) & & \\[2pt]
\hline
$[\mathrm{C}/\mathrm{H}]$ & \gradientci{-0.054}{0.071}{0.000} & \gradientci{0.013}{0.052}{0.018} & \gradientci{15.03}{6.85}{0.50} & \gradientci{-0.554}{0.026}{0.325} & $0.000$ & 304 & 62 \\[1pt]
$[\mathrm{O}/\mathrm{H}]$ & \gradientci{-0.085}{0.045}{0.012} & \gradientci{-0.005}{0.025}{0.018}\textsuperscript{\ensuremath{\dagger}} & \gradientci{10.78}{2.87}{1.23} & \gradientci{-0.350}{0.067}{0.201} & $0.000$ & 329 & 154 \\[1pt]
$[\mathrm{Na}/\mathrm{H}]$ & \gradientci{-0.137}{0.010}{0.047} & \gradientci{-0.034}{0.002}{0.019} & \gradientci{7.92}{0.04}{2.83} & \gradientci{0.143}{0.190}{0.009} & $0.096$ & 322 & 230 \\[1pt]
$[\mathrm{Mg}/\mathrm{H}]$ & \gradientci{-0.053}{0.065}{0.001} & \gradientci{0.023}{0.057}{0.011} & \gradientci{14.98}{7.07}{0.20} & \gradientci{-0.378}{0.015}{0.316} & $0.085$ & 330 & 67 \\[1pt]
$[\mathrm{Al}/\mathrm{H}]$ & \gradientci{-0.063}{0.043}{0.000} & \gradientci{0.048}{0.096}{0.012} & \gradientci{15.50}{7.62}{0.15} & \gradientci{-0.385}{0.017}{0.452} & $0.079$ & 217 & 30 \\[1pt]
$[\mathrm{Si}/\mathrm{H}]$ & \gradientci{-0.067}{0.013}{0.002} & \gradientci{0.029}{0.031}{0.017}\textsuperscript{\ensuremath{\dagger}} & \gradientci{14.90}{2.15}{0.29} & \gradientci{-0.509}{0.022}{0.072} & $0.070$ & 259 & 43 \\[1pt]
$[\mathrm{Ca}/\mathrm{H}]$ & \gradientci{-0.055}{0.026}{0.001} & \gradientci{0.011}{0.056}{0.009} & \gradientci{15.15}{7.26}{0.09} & \gradientci{-0.392}{0.013}{0.380} & $0.078$ & 334 & 58 \\[1pt]
$[\mathrm{Ti}/\mathrm{H}]$ & \gradientci{-0.059}{0.004}{0.003} & \gradientci{0.042}{0.014}{0.024}\textsuperscript{\ensuremath{\dagger}} & \gradientci{15.24}{0.26}{0.41} & \gradientci{-0.492}{0.026}{0.022} & $0.089$ & 334 & 56 \\[1pt]
$[\mathrm{Cr}/\mathrm{H}]$ & \gradientci{-0.068}{0.015}{0.002} & \gradientci{0.027}{0.077}{0.018}\textsuperscript{\ensuremath{\dagger}} & \gradientci{15.18}{7.20}{0.16} & \gradientci{-0.571}{0.022}{0.445} & $0.082$ & 335 & 56 \\[1pt]
$[\mathrm{Mn}/\mathrm{H}]$ & \gradientci{-0.141}{0.013}{0.054} & \gradientci{-0.043}{0.001}{0.035} & \gradientci{7.89}{0.08}{3.84} & \gradientci{0.018}{0.296}{0.008} & $0.106$ & 263 & 186 \\[1pt]
$[\mathrm{Fe}/\mathrm{H}]$ & \gradientci{-0.062}{0.003}{0.003} & \gradientci{0.034}{0.014}{0.029}\textsuperscript{\ensuremath{\dagger}} & \gradientci{15.28}{0.19}{0.44} & \gradientci{-0.484}{0.031}{0.020} & $0.120$ & 335 & 56 \\[1pt]
$[\mathrm{Co}/\mathrm{H}]$ & \gradientci{-0.103}{0.017}{0.018} & \gradientci{-0.007}{0.004}{0.006}\textsuperscript{\ensuremath{\dagger}} & \gradientci{7.90}{0.03}{0.02} & \gradientci{0.087}{0.023}{0.013} & $0.123$ & 181 & 103 \\[1pt]
$[\mathrm{Ni}/\mathrm{H}]$ & \gradientci{-0.118}{0.013}{0.021} & \gradientci{-0.033}{0.002}{0.006} & \gradientci{7.92}{0.03}{0.26} & \gradientci{-0.174}{0.030}{0.013} & $0.000$ & 262 & 171 \\[1pt]
$[\mathrm{Y}/\mathrm{H}]$ & \gradientci{-0.048}{0.006}{0.003} & \gradientci{0.008}{0.019}{0.019} & \gradientci{15.19}{0.63}{0.26} & \gradientci{-0.435}{0.024}{0.035} & $0.000$ & 335 & 56 \\[1pt]
$[\mathrm{Ba}/\mathrm{H}]$ & \gradientci{-0.004}{0.053}{0.024} & \gradientci{-0.057}{0.008}{0.080}\textsuperscript{\ensuremath{\dagger}} & \gradientci{9.08}{0.36}{6.66} & \gradientci{0.347}{0.421}{0.033} & $0.229$ & 296 & 171 \\[1pt]
$[\mathrm{Eu}/\mathrm{H}]$ & \gradientci{-0.043}{0.004}{0.002} & \gradientci{0.031}{0.015}{0.025}\textsuperscript{\ensuremath{\dagger}} & \gradientci{15.28}{0.34}{0.43} & \gradientci{-0.516}{0.024}{0.021} & $0.096$ & 327 & 55 \\[1pt]
$[\mathrm{M}/\mathrm{H}]$ & \gradientci{-0.055}{0.005}{0.002} & \gradientci{0.027}{0.018}{0.028}\textsuperscript{\ensuremath{\dagger}} & \gradientci{15.01}{1.18}{0.62} & \gradientci{-0.459}{0.031}{0.033} & $0.087$ & 335 & 66 \\[1pt]
\hline
\end{tabular*}
\par\vspace{0.6em}
\begin{minipage}{\linewidth}
\scriptsize\raggedright
Notes. The fitted abundance $y$ is the ratio in each row. Linear model: $y=mR_{\mathrm{GC}}+b$. Continuous broken-linear model: $y=y_{\mathrm{b}}+m_i(R_{\mathrm{GC}}-R_{\mathrm{b}})$, with $i=1$ for $R_{\mathrm{GC}}\leq R_{\mathrm{b}}$ and $i=2$ otherwise.
\par\smallskip
Linear uncertainties are bootstrap sample standard deviations. Broken-linear values $v_{-\Delta_-}^{+\Delta_+}$ are offsets to the 16th and 84th bootstrap percentiles. We attempt $10^{4}$ bootstrap resamples and retain all valid fits with no clipping. A $\dagger$ identifies a slope whose bootstrap standard deviation exceeds five times the half-width of the 68\% confidence interval, indicating substantial tails in the bootstrap distribution.
\par\smallskip
Intrinsic scatter $s_{\mathrm{int}}$ is the extra vertical abundance dispersion beyond measurement uncertainties. Slopes are in dex kpc$^{-1}$; $b$, $y_{\mathrm{b}}$, and $s_{\mathrm{int}}$ in dex; $R_{\mathrm{b}}$ in kpc. $N_{\mathrm{stars}}$ is the number of stars fitted; $N_{\mathrm{break}}$ counts stars with observed $R_{\mathrm{GC}}>R_{\mathrm{b}}$.
\par\smallskip
All abundances are in NLTE except for C and Cr. 
\end{minipage}
\end{table}

\clearpage
\section{Catalog of atmospheric parameters and abundances}
\label{app:catalog}

The full catalog with all the derived atmospheric parameters and abundances for the target elements in our self-consistent NLTE analysis can be found in Table \ref{tab:Catalog}. The combined catalog for all stars in our sample is shown in Table \ref{tab:AveragedCatalog}.

\begin{table}[H]
\renewcommand{\arraystretch}{0.90}
\centering
\scriptsize
\setlength{\tabcolsep}{2.5pt}
\caption{Individual atmospheric parameters and abundances.}
\label{tab:Catalog}
\begin{tabular*}{\textwidth}{@{\extracolsep{\fill}} lrrllrrrrrrrrrrr}
\toprule
Name & MJD & $\varphi$ & Program & Inst. & $T_{\rm eff}$ & $\log g$ & $\rm{V}_{\rm{mic}}$ & $\rm{V}_{\rm{brd}}$ & RV & $[\mathrm{Fe/H}]$ & $[\mathrm{Mg/Fe}]$ & $[\mathrm{Ca/Fe}]$ & $[\mathrm{Si/Fe}]$ & \ldots & $[\mathrm{Eu/Fe}]$ \\
  & d &  &  &  & K & dex & km s$^{-1}$ & km s$^{-1}$ & km s$^{-1}$ & dex & dex & dex & dex & ... & dex \\
\midrule
AB Vel & 57366.27112 & 0.54 & 096.D-0635(A) & UVES & 5657 & 1.52 & 3.75 & 10.32 & 35.89 & 0.03 & -0.03 & 0.05 & 0.00 & ... & -0.09 \\
AD Cru & 55282.25245 & 0.03 & 084.B-0029(A) & FEROS & 6024 & 1.38 & 3.66 & 13.37 & -46.38 & -0.03 & -0.02 & 0.02 & -0.05 & ... & -0.24 \\
AD Cru & 55282.26878 & 0.03 & 084.B-0029(A) & FEROS & 6006 & 1.32 & 3.70 & 13.34 & -46.26 & -0.05 & -0.02 & 0.01 & -0.06 & ... & -0.22 \\
AD Cru & 55282.28188 & 0.03 & 084.B-0029(A) & FEROS & 5997 & 1.35 & 3.71 & 13.37 & -46.18 & -0.04 & -0.04 & 0.00 & -0.06 & ... & -0.22 \\
AH Vel & 53306.26199 & 0.34 & 074.D-0008(B) & FEROS & 6047 & 1.25 & 4.00 & 14.24 & 21.56 & -0.15 & 0.01 & 0.03 & -0.01 & ... & -0.23 \\
AH Vel & 55285.03576 & 0.45 & 084.B-0029(A) & FEROS & 5909 & 1.13 & 3.94 & 14.33 & 27.24 & -0.15 & 0.00 & 0.02 & -0.07 & ... & -0.16 \\
AH Vel & 55285.03704 & 0.45 & 084.B-0029(A) & FEROS & 5917 & 1.15 & 3.94 & 14.29 & 27.25 & -0.14 & -0.01 & 0.00 & -0.06 & ... & -0.17 \\
AH Vel & 55285.03833 & 0.45 & 084.B-0029(A) & FEROS & 5896 & 1.12 & 3.94 & 14.27 & 27.26 & -0.16 & -0.01 & 0.02 & -0.06 & ... & -0.16 \\
AH Vel & 56606.17654 & 0.99 & 190.D-0237(E) & HARPS & 6381 & 1.37 & 4.57 & 15.99 & 20.00 & -0.17 & 0.03 & 0.04 & -0.01 & ... & -0.16 \\
AH Vel & 56910.39981 & 0.96 & 190.D-0237(F) & HARPS & 6315 & 1.37 & 4.54 & 15.90 & 22.42 & -0.15 & -0.01 & 0.00 & -0.04 & ... & -0.20 \\
AT Pup & 55284.02096 & 0.04 & 084.B-0029(A) & FEROS & 6371 & 1.04 & 4.05 & 15.55 & 9.01 & -0.26 & 0.07 & 0.09 & 0.02 & ... & -0.27 \\
AV Sgr & 53615.03183 & 0.59 & 075.D-0676(A) & FEROS & 4083 & 2.06 & 2.13 & 22.33 & 27.82 & 0.43 & 0.14 & 0.09 & - & ... & 0.49 \\
AV Sgr & 56136.16400 & 0.28 & 089.D-0767(C) & UVES & 5117 & 0.97 & 3.67 & 10.07 & 19.95 & 0.19 & 0.09 & 0.02 & 0.11 & ... & -0.19 \\
AV Sgr & 56136.17444 & 0.28 & 089.D-0767(C) & UVES & 5114 & 0.98 & 3.68 & 10.09 & 20.14 & 0.15 & 0.11 & 0.05 & 0.13 & ... & -0.17 \\
AV Sgr & 56152.07738 & 0.31 & 089.D-0767(C) & UVES & 5068 & 1.00 & 3.78 & 10.63 & 22.69 & 0.15 & 0.16 & 0.04 & 0.14 & ... & -0.16 \\
AV Sgr & 56168.04459 & 0.35 & 089.D-0767(C) & UVES & 5018 & 1.02 & 3.92 & 11.55 & 27.10 & 0.12 & 0.20 & 0.03 & 0.16 & ... & -0.13 \\
AX Cir & 53403.26673 & 0.24 & 074.D-0008(B) & FEROS & 5699 & 1.42 & 3.26 & 14.36 & -39.00 & -0.16 & 0.01 & 0.07 & 0.01 & ... & -0.16 \\
AX Cir & 55283.42180 & 0.76 & 084.B-0029(A) & FEROS & 5781 & 1.78 & 3.93 & 20.00 & -9.00 & -0.13 & 0.08 & 0.02 & -0.04 & ... & -0.13 \\
AX Cir & 56876.98997 & 0.95 & 190.D-0237(E) & HARPS & 6114 & 1.48 & 3.88 & 17.05 & -27.06 & -0.18 & -0.03 & 0.03 & -0.10 & ... & -0.21 \\
AX Cir & 56877.99425 & 0.14 & 190.D-0237(E) & HARPS & 5760 & 1.28 & 3.32 & 14.80 & -23.86 & -0.19 & -0.02 & 0.06 & -0.09 & ... & -0.21 \\
... & ... & ... & ... & ... & ... & ... & ... & ... & ... & ... & ... & ... & ... & ... & ... \\
DO Ori & 57334.28209 & 0.83 & 096.D-0635(A) & UVES & 6614 & 3.00 & 4.04 & 21.24 & 90.16 & -0.34 & 0.23 & -0.13 & -0.02 & ... & -0.22 \\
... & ... & ... & ... & ... & ... & ... & ... & ... & ... & ... & ... & ... & ... & ... & ... \\
EF Tau & 57307.32032 & 0.13 & 096.D-0635(A) & UVES & 6037 & 1.22 & 3.34 & 15.37 & -16.81 & -0.92 & 0.23 & 0.16 & 0.04 & ... & 0.00 \\
... & ... & ... & ... & ... & ... & ... & ... & ... & ... & ... & ... & ... & ... & ... & ... \\
BF Ser & 60065.09381 & 0.25 & 111.24YL.001 & UVES & 7030 & 2.99 & 3.16 & 0.84 & -163.66 & -1.80 & 0.39 & 0.28 & 0.32 & ... & 0.87 \\
... & ... & ... & ... & ... & ... & ... & ... & ... & ... & ... & ... & ... & ... & ... & ... \\
zet Gem & 53037.16424 & 0.55 & 072.D-0419(B) & HARPS & 5415 & 1.26 & 4.20 & 15.56 & 21.77 & 0.01 & 0.07 & 0.06 & 0.01 & ... & -0.22 \\
\bottomrule
\end{tabular*}
\tablefoot{This table is available in machine-readable form in the online article.}
\end{table}

\begin{table}[H]
\renewcommand{\arraystretch}{0.90}
\centering
\scriptsize
\setlength{\tabcolsep}{3.3pt}
\caption{Extract of the averaged stellar abundance catalog.}
\label{tab:AveragedCatalog}
\begin{tabular*}{\textwidth}{@{\extracolsep{\fill}} llrrrrrrrr}
\toprule
Name & Type & RA & Dec & $N_{\rm spec}$ & $[\mathrm{Fe/H}]$ & $[\mathrm{Mg/Fe}]$ & $[\mathrm{Ca/Fe}]$ & \ldots & $[\mathrm{Eu/Fe}]$ \\
 &  & deg & deg &   & dex & dex & dex &  & dex \\
\midrule
AB Vel & DCEP & 154.967074 & -56.309776 & 1 & $0.03 \pm 0.06$ & $-0.03 \pm 0.05$ & $0.05 \pm 0.07$ & \ldots & $-0.09 \pm 0.10$ \\
AD Cru & DCEP & 183.248564 & -62.096823 & 3 & $-0.04 \pm 0.04$ & $-0.03 \pm 0.03$ & $0.01 \pm 0.04$ & \ldots & $-0.23 \pm 0.06$ \\
AH Vel & DCEP & 122.999834 & -46.644319 & 6 & $-0.15 \pm 0.03$ & $0.00 \pm 0.02$ & $0.02 \pm 0.03$ & \ldots & $-0.18 \pm 0.04$ \\
AT Pup & DCEP & 123.093321 & -36.943742 & 1 & $-0.26 \pm 0.06$ & $0.07 \pm 0.05$ & $0.09 \pm 0.07$ & \ldots & $-0.27 \pm 0.10$ \\
AV Sgr & DCEP & 271.203275 & -22.732415 & 5 & $0.16 \pm 0.03$ & $0.14 \pm 0.03$ & $0.03 \pm 0.03$ & \ldots & $-0.05 \pm 0.05$ \\
AX Cir & DCEP & 223.146839 & -63.809863 & 5 & $-0.17 \pm 0.03$ & $0.01 \pm 0.02$ & $0.04 \pm 0.03$ & \ldots & $-0.19 \pm 0.04$ \\
AY Cen & DCEP & 171.273919 & -60.734612 & 1 & $-0.07 \pm 0.06$ & $0.00 \pm 0.05$ & $0.02 \pm 0.07$ & \ldots & $-0.22 \pm 0.10$ \\
AZ Cen & DCEP & 171.304037 & -61.369148 & 1 & $-0.14 \pm 0.06$ & $0.03 \pm 0.05$ & $0.03 \pm 0.07$ & \ldots & $-0.22 \pm 0.10$ \\
BE Mon & DCEP & 100.023257 & 7.605835 & 1 & $-0.09 \pm 0.06$ & $0.02 \pm 0.05$ & $0.03 \pm 0.07$ & \ldots & $-0.18 \pm 0.10$ \\
BH Vel & DCEP & 127.452113 & -41.223776 & 2 & $-0.17 \pm 0.04$ & $0.02 \pm 0.03$ & $0.06 \pm 0.05$ & \ldots & $-0.12 \pm 0.07$ \\
BK Cen & DCEP & 177.316647 & -63.078576 & 2 & $0.04 \pm 0.04$ & $-0.03 \pm 0.03$ & $0.01 \pm 0.05$ & \ldots & $-0.15 \pm 0.07$ \\
BP Cir & DCEP & 221.674867 & -61.461959 & 2 & $-0.17 \pm 0.05$ & $0.00 \pm 0.04$ & $0.03 \pm 0.05$ & \ldots & $-0.23 \pm 0.07$ \\
BR Vul & DCEP & 296.646600 & 22.889753 & 7 & $0.04 \pm 0.02$ & $-0.04 \pm 0.02$ & $0.03 \pm 0.03$ & \ldots & $-0.17 \pm 0.04$ \\
bet Dor & DCEP & 83.406311 & -62.489769 & 41 & $-0.10 \pm 0.01$ & $0.03 \pm 0.01$ & $0.06 \pm 0.01$ & \ldots & $-0.21 \pm 0.02$ \\
CC Car & DCEP & 163.611890 & -61.347099 & 3 & $0.01 \pm 0.04$ & $-0.05 \pm 0.03$ & $0.03 \pm 0.04$ & \ldots & $-0.20 \pm 0.06$ \\
CF Car & DCEP & 164.770433 & -60.278842 & 4 & $0.05 \pm 0.03$ & $0.01 \pm 0.02$ & $0.05 \pm 0.03$ & \ldots & $-0.17 \pm 0.05$ \\
CK Pup & DCEP & 111.828684 & -25.785354 & 2 & $-0.14 \pm 0.06$ & $0.02 \pm 0.03$ & $0.00 \pm 0.05$ & \ldots & - \\
CM Pup & DCEP & 112.676617 & -20.451986 & 1 & $-0.42 \pm 0.06$ & $0.05 \pm 0.04$ & $0.07 \pm 0.07$ & \ldots & $-0.04 \pm 0.10$ \\
CO Vel & DCEP & 127.422122 & -40.333882 & 2 & $-0.07 \pm 0.04$ & $-0.01 \pm 0.03$ & $0.02 \pm 0.05$ & \ldots & $-0.13 \pm 0.07$ \\
CP Vel & DCEP & 132.043641 & -48.659152 & 1 & $-0.17 \pm 0.06$ & $0.06 \pm 0.04$ & $0.03 \pm 0.07$ & \ldots & $-0.17 \pm 0.10$ \\
\ldots & \ldots & \ldots & \ldots & \ldots & \ldots & \ldots & \ldots & \ldots & \ldots \\
DO Ori & RRAB & 90.459080 & 15.401440 & 1 & $-0.34 \pm 0.06$ & $0.23 \pm 0.04$ & $-0.13 \pm 0.07$ & \ldots & $-0.22 \pm 0.17$ \\
\ldots & \ldots & \ldots & \ldots & \ldots & \ldots & \ldots & \ldots & \ldots & \ldots \\
EF Tau & T2CEP & 85.882478 & 19.420019 & 1 & $-0.92 \pm 0.06$ & $0.23 \pm 0.04$ & $0.16 \pm 0.07$ & \ldots & $0.00 \pm 0.10$ \\
\ldots & \ldots & \ldots & \ldots & \ldots & \ldots & \ldots & \ldots & \ldots & \ldots \\
BF Ser & ACEP & 229.118680 & 16.444301 & 3 & $-1.79 \pm 0.04$ & $0.36 \pm 0.02$ & $0.27 \pm 0.04$ & \ldots & $0.87 \pm 0.19$ \\
\ldots & \ldots & \ldots & \ldots & \ldots & \ldots & \ldots & \ldots & \ldots & \ldots \\
zet Gem & DCEP & 106.027177 & 20.570294 & 35 & $-0.01 \pm 0.01$ & $0.03 \pm 0.01$ & $0.07 \pm 0.01$ & \ldots & $-0.22 \pm 0.02$ \\
\bottomrule
\end{tabular*}

\tablefoot{Each row represents one star. Abundances are inverse variance weighted means, in dex relative to the adopted solar reference; the quoted uncertainties are the formal propagated errors of those means, assuming independent measurements. C and Cr are determined in LTE, while Fe and the other elements are determined in NLTE. This table is available in machine-readable form in the online article.
 }
\end{table}

\end{appendix}

\end{document}